\documentclass[aps,prl,twocolumn,longbibliography]{revtex4-1}
\usepackage{bbm}
\usepackage{graphicx}
\usepackage{dcolumn}
\usepackage{bm}
\usepackage{subfigure}
\usepackage{amsmath}
\usepackage{feynmf}
\usepackage{hyperref}
\usepackage{CJK}
 \usepackage{amssymb}

\usepackage{attachfile}

\newcommand{\bk}{\boldsymbol k}

\newcommand{\bq}{\boldsymbol q}

\newcommand{\bd}{\boldsymbol d}

\newcommand{\bM}{\boldsymbol M}
\newcommand{\bv}{\boldsymbol v}
\newcommand{\bOmega}{\boldsymbol{\Omega}}
\newcommand{\bS}{\boldsymbol{S}}
\newcommand{\bsigma}{\boldsymbol{\sigma}}
\newcommand{\bA}{\boldsymbol{A}}

\usepackage{braket}
\usepackage{esint}
\usepackage{times}

\begin{document}

\title{Berry-dipole Semimetals}

\author{Zheng-Yang Zhuang}
\affiliation{Guangdong Provincial Key Laboratory of Magnetoelectric Physics and Devices,
State Key Laboratory of Optoelectronic Materials and Technologies,
School of Physics, Sun Yat-sen University, Guangzhou 510275, China}

\author{Chaoyi Zhang}
\affiliation{Guangdong Provincial Key Laboratory of Magnetoelectric Physics and Devices,
State Key Laboratory of Optoelectronic Materials and Technologies,
School of Physics, Sun Yat-sen University, Guangzhou 510275, China}

\author{Xiao-Jiao Wang}
\affiliation{Guangdong Provincial Key Laboratory of Magnetoelectric Physics and Devices,
State Key Laboratory of Optoelectronic Materials and Technologies,
School of Physics, Sun Yat-sen University, Guangzhou 510275, China}

\author{Zhongbo Yan}
\email{yanzhb5@mail.sysu.edu.cn}
\affiliation{Guangdong Provincial Key Laboratory of Magnetoelectric Physics and Devices,
State Key Laboratory of Optoelectronic Materials and Technologies,
School of Physics, Sun Yat-sen University, Guangzhou 510275, China}

\date{\today}

\begin{abstract}
We introduce ``Berry-dipole semimetals'', whose band degeneracies are characterized by quantized
Berry dipoles. Through a two-band model  constructed by Hopf map, we reveal that the Berry-dipole semimetals
display a multitude of salient properties distinct from
other topological semimetals. On the boundary,  we find that
the first-order Berry-dipole semimetal harbors anomalous paired Fermi arcs with the same spin polarization,  even
though the layer Chern number is zero, and the second-order Berry-dipole semimetal hosts dispersionless hinge arcs. In
the bulk, we find that the low-energy Berry-dipole Hamiltonian near the band node has
a quadratic energy dispersion and peculiar Berry curvature, which give rise
to rather unique characteristics in the intrinsic anomalous Hall effect,
orbital magnetization and Landau levels.
Our study shows that Berry-dipole semimetals
are a class of topological gapless phases supporting rich intriguing physics.
\end{abstract}

\maketitle

Over the past decade,  topological semimetals in three dimensions (3D),
including nodal-point and nodal-line semimetals,  have been intensively and
extensively investigated both in theory
and in experiment~\cite{Yan2016review,Burkov2016,Armitage2018RMP,Lv2021RMP,Nagaosa2020,Hasan2021}.
The Weyl semimetal (WSM) is a paradigmatic nodal-point semimetal attracting particular
interest~\cite{wan2011,Xu2011weyl,burkov2011,Weng2015weyl,Lv2015weyl,Xu2015weyl,Huang2015weyl,Zhang2016weyl}. The band degeneracy in WSMs, known as Weyl node (or point),
is a singularity of the Berry curvature and
a topological object whose stability does not need symmetry protection~\cite{Volovik2001review}.
The low-energy
Hamiltonian near the Weyl node, which resembles the
relativistic Weyl Hamiltonian, gives rise to a linear
energy dispersion and a Berry curvature structure whose
momentum dependence is reminiscent of the field distribution of a magnetic monopole in real
space~\cite{Berry1984,volovik1987zeros}. Because of the relativistic linear dispersion and monopole-like Berry curvature,
WSMs carry a diversity of peculiar properties, such as
open Fermi arcs~\cite{wan2011} and chiral Landau levels~\cite{Yuan2018,Jia2019chiral}, and serve as a fertile playground
to investigate exotic electromagnetic field responses~\cite{Son2012chiral,Zyuzin2012weyl,Vazifeh2013weyl,Nandy2017planar,deJuan2017,Yuan2020,Sekine2021axion} and interaction-driven phases~\cite{Cho2012,wang2013a,Roy2017weyl}.

Since the Weyl node acts as a charge of Berry flux, a natural question
is whether one can extend the concept of charge to dipole. Naively, when there are two Weyl nodes with opposite charges,
one may expect that they naturally consist of a Berry dipole.
However, such a picture is rather trivial since, in WSMs,  the Weyl nodes with
opposite charges always show up in pairs~\cite{Nielsen1981nogo,Nielsen1981a,Nielsen1981b}. More importantly,
the physics remains dominated by the Weyl Hamiltonian,
without the need of a dipole picture. Interestingly, a dipole
can also be point-like, and the local and global
properties of a point dipole are completely different from a point charge.
Therefore, an extension to point Berry dipole
would be nontrivial.

\begin{figure}[t]
	\centering
	\subfigure{
		\includegraphics[scale=0.45]{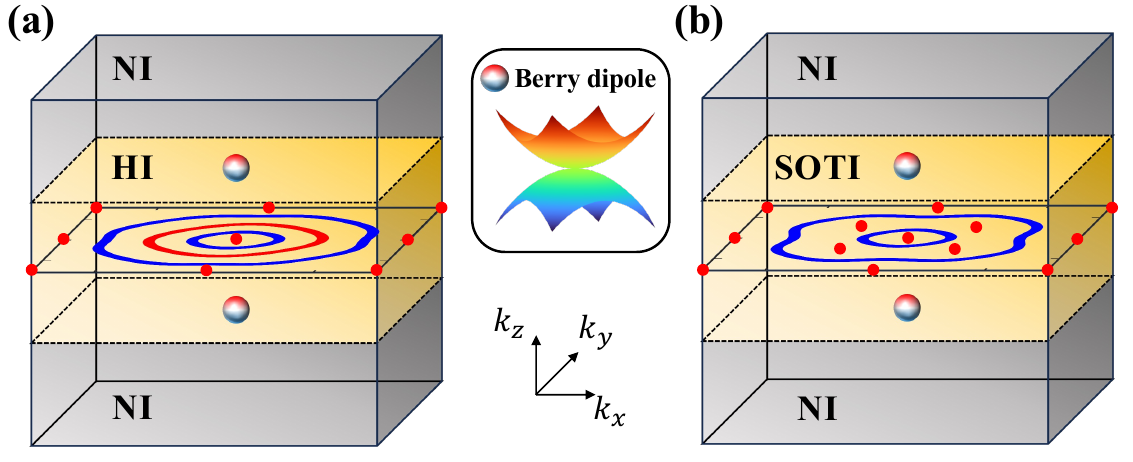}}
	\caption{Schematic of a first-order Berry-dipole semimetal [(a)] and  a second-order Berry-dipole semimetal [(b)]. The quadratic energy dispersion near the Berry-dipole node is sketched in the middle. In (a) and (b), the $k_{z}$-planes in yellow correspond to 2D Hopf insulators (HIs) and second-order topological insulators (SOTIs), respectively, and the $k_{z}$-planes in gray correspond to normal insulators (NIs).
The blue and red lines (dots) refer to the solutions of $d_{z}(\bk)=0$ and $d_{x}(\bk)=d_{y}(\bk)=0$ at the $k_{z}=0$ plane, respectively, which illustrate why the yellow planes are topologically nontrivial.}
	\label{Fig1}
\end{figure}

A point dipole is formed by two opposite point charges that are infinitely close.
If there is no symmetry protection, however, the collision
of two Weyl nodes with opposite charges will cause an annihilation
of them. Recent works have shown that if two Weyl nodes are related by a mirror symmetry,
their annihilation is avoided~\cite{Sun2018mirrorweyl}, and a point Berry dipole
will form when they meet at the mirror plane. Furthermore,
it was found that the critical point between two different
3D crystalline Hopf insulators is also a point Berry dipole~\cite{Nelson2021hopf}.
Inspired by the existence of point Berry dipole, in this work
we introduce the concept of ``Berry-dipole semimetal'' (BDSM).
Rather than as the critical point of two 3D crystalline Hopf insulators,
we find that, using the dimension-reduction picture, the Berry-dipole node can also show up at the
plane separating 2D Hopf-insulator/second-order-topological-insulator planes and normal-insulator planes in the 3D Brillouin zone
(see the illustration in Figs.\ref{Fig1}(a) and \ref{Fig1}(b)), just like that the Weyl node
sits at the plane separating 2D Chern-insulator planes and normal-insulator planes~\cite{yang2011}.
With this observation in mind, we use the Hopf map to construct a two-band model
that can realize both first-order and second-order BDSMs.
As expected, we find that the boundary and bulk states of the BDSMs exhibit many salient properties
distinct from other topological semimetals.

{\it General theory.---}The Hopf map refers to a map from $3$-sphere
($S^{3}$) to 2-sphere ($S^{2}$). Mathematically, this map can be described
by a map from a two-component complex spinor to a three-component real vector.
Denoting the complex spinor as $\zeta=(\zeta_{1},\zeta_{2})^{T}$ and the real vector
as $\bd=(d_{x},d_{y},d_{z})$, they are related by the formula,
$\bd=\zeta^{\dag}\boldsymbol{\sigma}\zeta$, where $\boldsymbol{\sigma}=(\sigma_{x},\sigma_{y},\sigma_{z})$
are the Pauli matrices. To be more explicit,
\begin{eqnarray}
d_{x}=2\text{Re}(\zeta_{1}^{*}\zeta_{2}),\,
d_{y}=2\text{Im}(\zeta_{1}^{*}\zeta_{2}),\,
d_{z}=|\zeta_{1}|^{2}-|\zeta_{2}|^{2}.
\end{eqnarray}
Using $d_{i}$ to construct a two-band Hamiltonian, $\mathcal{H}(\bk)=\bd(\bk)\cdot\boldsymbol{\sigma}$,
it has been demonstrated that diverse distinctive topological phases can be realized by appropriately choosing
the complex spinor. Notable examples include 3D Hopf insulators~\cite{Moore2008hopf,
deng2013hopf,Aris2021hopf,Zhu2021hopf,Zhu2023spinhopf,Lim2023hopf,Wang2023hopf}, 3D nodal-link
semimetals~\cite{Yan2017link,Chen2017link,Ezawa2017hopf}, 3D dipolar WSMs~\cite{Tyner2024dipolar}, 2D second-order
topological insulators/superconductors~\cite{Yan2019hosca,Niu2021hopf,Wu2023higher}, and Hopf defects harboring
Majorana zero modes in 2D trivial superconductors~\cite{Yan2017hopf}. In this work, we choose
\begin{eqnarray}
\zeta_{1}(\bk)&=&\lambda(\sin k_{x}+i\sin k_{y}),\nonumber\\
\zeta_{2}(\bk)&=&\delta(\cos k_{x}-\cos k_{y})+i(M-t\sum_{i=x,y,z}\cos k_{i}).
\end{eqnarray}
As will be shown below, when $\delta=0$, the two-band model can realize
both first-order and second-order BDSMs, depending on whether $\delta$ is zero or finite.

{\it First-order BDSM.---}We first consider the case with $\delta=0$.
The components of the Hamiltonian are then given by
\begin{eqnarray}
d_{x}(\bk)&=&2\lambda M(\bk)\sin k_{y},\,d_{y}(\bk)=2\lambda M(\bk)\sin k_{x},\nonumber\\
d_{z}(\bk)&=&\lambda^{2}\sin^{2}k_{x}+\lambda^{2}\sin^{2}k_{y}-M^{2}(\bk).\label{FH}
\end{eqnarray}
Here we have introduced $M(\bk)=(M-t\sum_{i=x,y,z}\cos k_{i})$ to shorten the notation,
and the lattice constants are set to unit throughout this work for notational simplicity.
The energy spectra of the Hamiltonian are
\begin{eqnarray}
E_{\pm}(\bk)=\pm[\lambda^{2}(\sin^{2}k_{x}+\sin^{2}k_{y})+M^{2}(\bk)].
\end{eqnarray}
Apparently, the energy bands harbor nodes if $M(\bk)$ has zeros on the four high symmetry
$k_{z}$ axes passing $(k_{x},k_{y})=(0/\pi,0/\pi)$. Without loss of generality, we consider
$t<M<3t$, then there are two band nodes  at $\bk_{n,\chi}=\chi(0,0,\arccos (M/t-2))$, where
$\chi=\pm$. To know the properties of the band nodes, we determine the low-energy Hamiltonians around the two band nodes, which
read
\begin{eqnarray}
\mathcal{H}_{\chi}(\bq)&=&2\chi vv_{z}q_{z}q_{y}\sigma_{x}+2\chi vv_{z}q_{z} q_{x}\sigma_{y}\nonumber\\
&&+[v^{2}(q_{x}^{2}+q_{y}^{2})-v_{z}^{2}q_{z}^{2}]\sigma_{z},\label{BPHamiltonian}
\end{eqnarray}
where $v=\lambda$, $v_{z}=t\sqrt{1-(M/t-2)^{2}}$ and $\bq=(q_{x},q_{y},q_{z})$ denotes
the momentum measured from the corresponding node. The low-energy spectra are
\begin{eqnarray}
E_{\chi,\pm}(\bq)=\pm[v^{2}(q_{x}^{2}+q_{y}^{2})+v_{z}^{2}q_{z}^{2}].
\end{eqnarray}
Obviously, the low-energy spectra are quadratic in all directions,
which is distinct from the Weyl and Dirac semimetals
hosting a linear dispersion near the band nodes.

A well-known characteristic of the Weyl node is that an integral of the Berry curvature
over a closed surface enclosing it is quantized. The quantized value corresponds to
the first-class Chern number and labels the topological charge of the Weyl node.
For the low-energy Hamiltonians in Eq.(\ref{BPHamiltonian}), their Berry curvatures are
given by~\cite{Zhuang2023NLHE}
\begin{eqnarray}
\boldsymbol{\Omega}_{\chi}^{(c)}(\bq)=-\boldsymbol{\Omega}_{\chi}^{(v)}(\bq)
=-\frac{2v^{2}v_{z}^{2}q_{z}\bq}{[v^{2}(q_{x}^{2}+q_{y}^{2})+v_{z}^{2}q_{z}^{2}]^{2}}.\label{BC}
\end{eqnarray}
The superscript ``$c/v$'' stands for conduction/valence band.
The form of the Berry curvatures resembles
the field distribution of a $z$-direction point dipole,
therefore, this kind of band node is dubbed Berry-dipole node~\cite{Nelson2021hopf,Graf2023Hopf}.

A notable property of the dipole Berry curvatures in Eq.(\ref{BC}) is that
an integral  over a closed surface enclosing the dipole is equal to zero,
while an integral over the upper half ($q_{z}>0$) or lower half ($q_{z}<0$) of the closed surface leads to
a nonzero but opposite quantized value~\cite{Nelson2021hopf,Zhuang2023NLHE}, i.e.,
\begin{eqnarray}
\frac{1}{2\pi}\int_{q_{z}\gtrless0}\boldsymbol{\Omega}_{\chi}^{(c)}(\bq)\cdot d\mathbf{S}=\mp1.\label{dipole}
\end{eqnarray}
The above result is a reflection of the fact that the Berry dipole is formed by two opposite and
infinitely close Berry charges.  From Eq.(\ref{BC}), one can also see that the two low-energy Berry-dipole Hamiltonians
lead to the same Berry curvature, which means that the two Berry dipoles are the same.
This fact suggests that the net Berry dipole does not need to vanish like
the net Berry charge in WSMs.

\begin{figure}[t]
	\centering
	\subfigure{
		\includegraphics[scale=0.45]{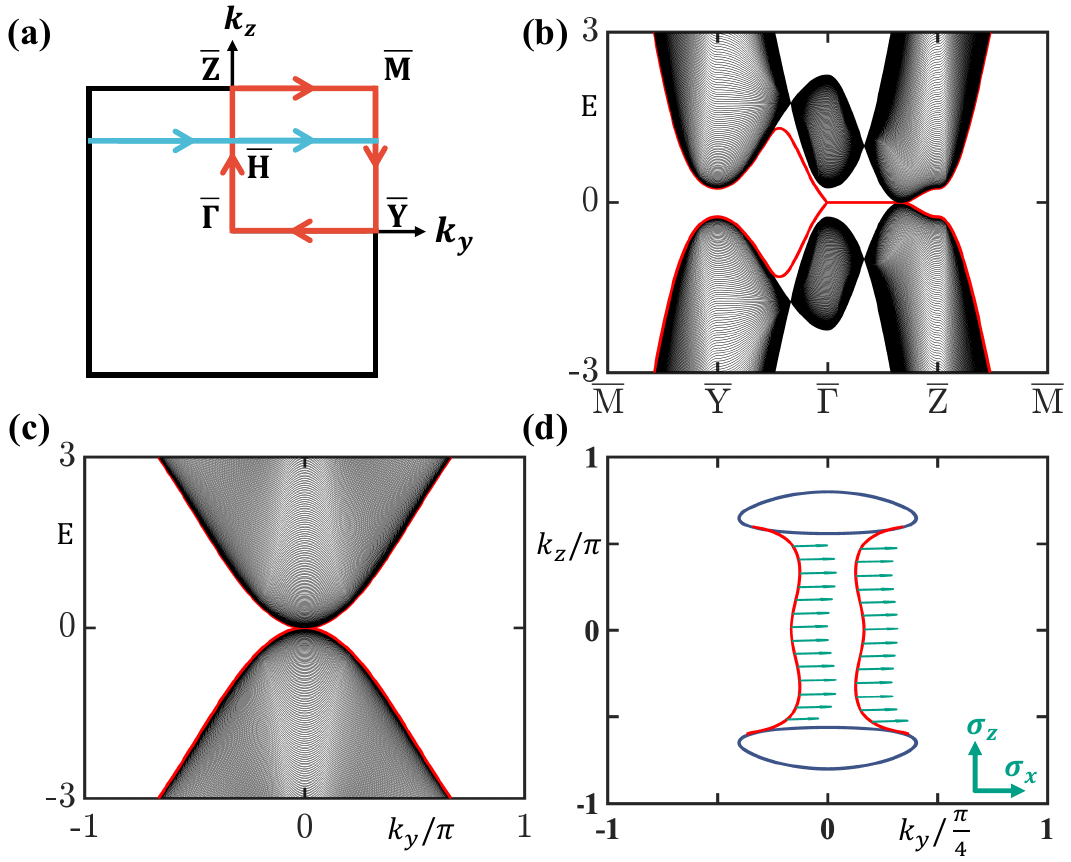}}
	\caption{(a) High symmetry points in the surface Brillouin zone.  $\rm \overline{H}$ refers to the location where one Berry-dipole node is projected. The energy spectra along the two paths in orange and light blue are plotted in (b) and (c). (d) Fermi arcs (red lines) and the projection of bulk Fermi surface (dark blue rings) on the right $x$-normal surface. The chemical potential is chosen to be $\mu=0.05$. The green arrows denote the spin polarizations on the Fermi arcs. Common parameters are $M=1.5$, $\lambda=t=1$ and $\delta=0$.}
	\label{Fig2}
\end{figure}

In WSMs, another well-known characteristic is the open Fermi arcs (isoenergy contour of the surface states' energy spectrum)
connecting the projections of the bulk Weyl nodes in the surface
Brillouin zone. A popular picture adopted to understand the origin of
Fermi arcs in WSMs is that the Hamiltonians on the 2D momentum planes lying between the Weyl nodes
describe Chern insulators with chiral edge states~\cite{yang2011}. For here the BDSM,
we find that the Chern number in each 2D plane between the two Berry-dipole nodes is zero.
Nevertheless, we find that, these 2D planes are topologically nontrivial and support a pair of gapless edge states, which
form a pair of Fermi arcs on each side surface, as shown
in Figs.\ref{Fig2}(a-d).  While the paired
feature of Fermi arcs resembles the scenario in Dirac semimetals (DSMs)~\cite{wang2012dirac,wang2013three}, the two systems
are essentially different.  In DSMs, a pair of Fermi arcs will
carry opposite spin textures as they are related by time-reversal symmetry (TRS). In contrast,
here the BDSM lacks TRS, and we find that the pair of Fermi arcs carry
the same spin polarization, as shown in Fig.\ref{Fig2}(d).

How to understand the counterintuitive existence of a pair of Fermi arcs when the layer
Chern number is zero and the TRS is absent?
We find that the two-band Hamiltonian at a given $k_{z}$ plane between the two Berry-dipole nodes
in fact describes a 2D two-band Hopf insulator~\cite{Wang2015hopf}, a class of fixed-band topological insulators beyond
the ten-fold way classification just similar to its 3D counterpart~\cite{Moore2008hopf,Schnyder2008,kitaev2009periodic,Ryu2010}.
Here we provide an intuitive picture
to understand why these 2D layer Hamiltonians with a fixed $k_{z}$ are topologically nontrivial and host
a pair of gapless edge states. To start, it is instructive to note the following three facts. First,
a system is topologically nontrivial as long as it cannot be adiabatically
transformed to an atomic insulator without the close of bulk or boundary energy gap.
Second, the first-order topology can change only when the bulk energy gap
gets closed. Third, the bulk energy gap of the BDSM Hamiltonian can only get closed
when $d_{x}(\bk)$, $d_{y}(\bk)$ and $d_{z}(\bk)$ simultaneously vanish
at some points of the Brillouin zone. These three facts
imply that the zero-value contours of $d_{x,y,z}(\bk)$ determine
whether it is topologically nontrivial or not.  In Fig.\ref{Fig1}, we have plotted the contours
satisfying $d_{z}(\bk)=0$ and the ones simultaneously satisfying $d_{x}(\bk)=0$
and $d_{y}(\bk)=0$  at the $k_{z}=0$ plane. Apparently, if the latter ones are fixed,
the former ones cannot adiabatically be made vanishing without the close of
bulk energy gap. This suggests that the 2D Hamiltonian at the $k_{z}=0$ plane
must be topologically nontrivial as it cannot be adiabatically transformed to an atomic insulator (all zero-value
contours are absent) if one fixes two of the three $d_{i}(\bk)$ terms.
For convenience, we dub Hamiltonians with such zero-value contour configurations
as {\it conditionally obstructed Hamiltonians}.

Notably, we find that the dimension of the boundary states of these 2D layer Hamiltonians
between the two Berry-dipole nodes is the same as the dimension of the zero-value contours lying between
the zero-value contours satisfying  $d_{z}(\bk)=0$ which cause obstruction.
When the dimension is one, the boundary states are 1D propagating
edge states; When the dimension is zero, the boundary states
are 0D corner states as will be shown in the next section discussing second-order BDSM.
As the Chern number is zero, the 1D edge states must show up in pair and be counter-propagating like the scenario
in a quantum spin Hall insulator~\cite{Kane2005a,Kane2005b,Bernevig2006HgTe}. The pair of edge states for the series of 2D layer Hamiltonians
between the two Berry-dipole nodes consists of the pair of Fermi arcs of the 3D BDSM.

The existence, stability and spin polarization of the pair of edge states can further be understood via a
1D topological invariant. The lattice Hamiltonian in Eq.(\ref{FH}) has a $C_{4z}$ rotation
symmetry. Therefore, without loss of generality, we consider the $k_{y}=0$ line in a $k_{z}$ plane for an illustration. On this
high symmetry line, the Hamiltonian reduces as
\begin{eqnarray}
\mathcal{H}(k_{x},k_{z})&=&2\lambda \tilde{M}(k_{x},k_{z})\sin k_{x}\sigma_{y}\nonumber\\
&&+[\lambda^{2}\sin^{2}k_{x}-\tilde{M}^{2}(k_{x},k_{z})]\sigma_{z},
\end{eqnarray}
where $\tilde{M}(k_{x},k_{z})=M-t-t\cos k_{x}-t\cos k_{z}$. $\mathcal{H}(k_{x},k_{z})$
has chiral symmetry as $\{\sigma_{x},\mathcal{H}(k_{x},k_{z})\}=0$. Therefore, the topological boundary states on
the $x$-normal surfaces can be characterized by a $k_{z}$-dependent winding
number~\cite{Ryu2010}, $W(k_{z})$.  Since
the two terms are just the real and imaginary part of the complex function, $[\lambda \sin k_{x}+i\tilde{M}(k_{x},k_{z})]^{2}$, $W(k_{z})=2$
as long as $\tilde{M}(k_{x},k_{z})=0$ has a solution for $k_{x}\in(0,\pi)$. $W(k_{z})=2$ explains
the existence and stability of the edge-state spectrum crossing at $k_{y}=0$ even without TRS, as shown in Fig.\ref{Fig2}(b).
On the other hand, since the zero-energy boundary states are also the eigenstates of the chiral symmetry operator, the
spin polarizations of the two zero-energy states enforced by $W(k_{z})=2$ on the same $x$-normal surface will take the
same direction, leading to that the pair of Fermi arcs on a given side surface carries the same spin-polarization
at the neutrality condition.

\begin{figure}[t]
	\centering
	\subfigure{
		\includegraphics[scale=0.3]{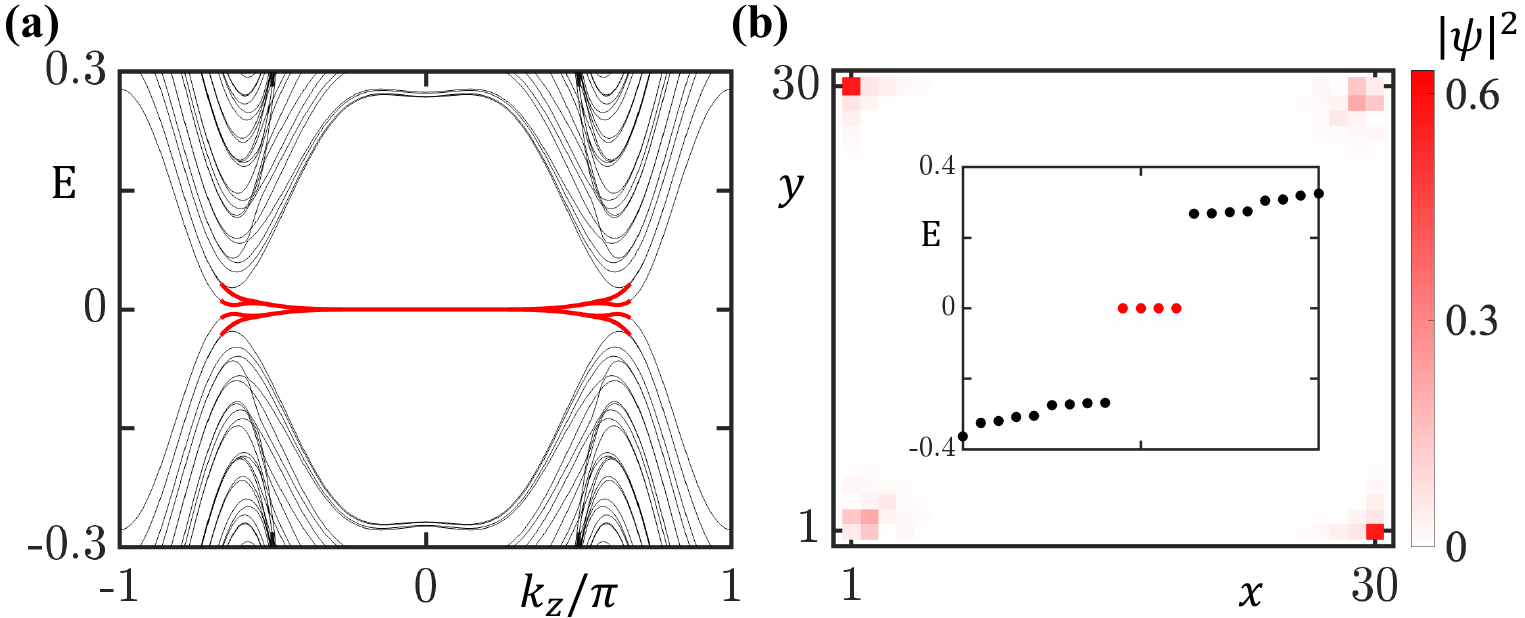}}
	\caption{(a) Energy spectra manifesting second-order topology. Periodic (open) boundary directions in the $z$ ($x$ and $y$) direction.
(b) The probability density profile of the four zero-energy bound states at $k_{z}=0$. Parameters are $M=1.5$, $\lambda=t=1$, and $\delta=0.2$.}
	\label{Fig3}
\end{figure}

{\it Second-order BDSM.---}When $\delta$ becomes finite, if one only keeps the leading order terms,
the low-energy Berry-dipole Hamiltonians retain their forms in Eq.(\ref{BPHamiltonian}).
Although the $\delta$ term has a negligible effect on the low-energy bulk physics,
its effect on the zero-value contour configurations is dramatic, and so as its effect
on the boundary states.  Once $\delta\neq0$,
the ring-shape zero-value contour lying between the two zero-value contours of $d_{z}(\bk)$
immediately becomes four separated points, as shown in Fig.\ref{Fig1}(b) (more detailed
discussions are given in Supplemental Material~\cite{supplemental}). This kind of
2D conditionally obstructed Hamiltonians correspond to second-order topological insulators/superconductors~\cite{Yan2019hosca}. Indeed,
considering open boundary conditions in $x$ and $y$ directions, we find one zero-energy bound state per corner when $k_{z}\in\{(\bk_{n,-})_{z},(\bk_{n,+})_{z}\}$, as shown in Fig.\ref{Fig3}. If one
restores the 3D perspective, these zero-energy bound states consist of four hinge Fermi arcs.
The above results suggest that the $\delta$ term  will gap out the surface Fermi arcs
 and lead to the presence of hinge Fermi arcs. In other words,
it renders a direct transition from a first-order BDSM
to a second-order one.

The bulk-hinge correspondence in the second-order BDSM shares more similarity
with a second-order DSM than a second-order WSM. In a second-order DSM,
hinge Fermi arcs can also be obtained by appropriately gapping out the helical surface Fermi arcs~\cite{Szabo2020HOTP}. In contrast,
the hinge Fermi arcs in a second-order WSM cannot directly be descended from the chiral surface Fermi arcs
as they are not gappable~\cite{Wang2020HOWSM,Ghorashi2020HOWSM,Wei2021,Luo2021}. In addition, if there are only two Weyl nodes in a second-order WSM,
the coexistence of surface and hinge Fermi arcs is inevitable.

{\it Intrinsic anomalous Hall effects and orbital magnetization.---}Besides the boundary states,
the quadratic energy dispersion and  dipole Berry curvature of the low-energy Hamiltonian can give rise to intriguing
measurable effects. Here we explore the intrinsic anomalous Hall effects and zero-field orbital magnetization
that are directly related to the Berry curvature. In 3D, the intrinsic anomalous Hall conductivity tensor has three independent components,
and they are connected to the Berry curvature according to the formula~\cite{Xiao2010BP} (hereafter we restore the reduced plank constant)
\begin{eqnarray}
\sigma_{ij}=\frac{e^{2}}{\hbar}\sum_{\chi}\sum_{n}\int\frac{d^{3}q}{(2\pi)^{3}}\epsilon_{ijk}\Omega_{\chi,k}^{(n)}f(E_{\chi,n}),
\end{eqnarray}
where $n$ is the band index and $f(E_{\chi,n})$ is the Fermi-Dirac distribution function. Considering
the zero-temperature limit and a small positive $\mu$ so that the low-energy Hamiltonian gives a very accurate
description, we find $\sigma_{xz}(\mu)=\sigma_{yz}(\mu)=0$, and~\cite{supplemental}
\begin{eqnarray}
\sigma_{xy}(\mu)=-\frac{2e^{2}}{\hbar}\frac{\sqrt{\mu}}{3\pi^{2} v_{z}}=-\frac{2e^{2}}{h}\frac{D_{F,z}}{3\pi},\label{Hall}
\end{eqnarray}
where $D_{F,z}=2\sqrt{\mu}/v_{z}$ has the geometric interpretation as the $z$-axis diameter of
the elliptic Fermi surface. For the two-band Hamiltonian, the orbital magnetic moment is simply
related to the Berry curvature~\cite{Xiao2007MM}, i.e.,
$\mathbf{m}_{\chi}(\bq)=(e/\hbar)E_{\chi,+}(\bq)\boldsymbol{\Omega}_{\chi}^{(c)}(\bq)$.
Notably, we find $m_{\chi,z}(0,0,q_{z})$ is a constant and can be expressed as~\cite{supplemental}
\begin{eqnarray}
m_{\chi,z}(0,0,q_{z})=\frac{-2ev^{2}}{\hbar}=-2\frac{e\hbar}{2M_{x(y)}^{*}}=-2\mu_{B}^{*},\label{OMM}
\end{eqnarray}
where $M_{x(y)}^{*}\equiv\hbar^{2}[\partial^{2}E_{\chi,+}/\partial q_{x(y)}^{2}]^{-1}=\hbar^{2}/2v^{2}$,  is the $xy$-plane effective mass of the conduction-band electrons. We emphasize that this result is different from
that of massive Dirac fermions at the band edge where the factor $2$ before the
effective Bohr magneton $\mu_{B}^{*}$ is absent~\cite{Xiao2007MM}.
The zero-field orbital magnetization is given by~\cite{Xiao2005OM}
\begin{eqnarray}
\mathbf{M}(\mu)&=&\sum_{\chi}\int^{\mu}\frac{d^{3}\bq}{(2\pi)^{3}}[\mathbf{m}_{\chi}(\bq)+\frac{e\boldsymbol{\Omega}_{\chi}^{(c)}(\bq)}{\hbar}(\mu-E_{\chi,+}(\bq))]\nonumber\\
&=&-\frac{2e}{\hbar}\frac{\mu^{3/2}}{3\pi^{2}v_{z}}\hat{z}=-2n_{T}\mu^{*}_{B}\hat{z}=\frac{\mu}{e}\sigma_{xy}(\mu)\hat{z},\label{OM}
\end{eqnarray}
where $n=\mu^{3/2}/(3\pi^{2}v^{2}v_{z})$ is the total electron density in the weakly doped regime. The result indicates that all electrons contribute the same magnetization.

\begin{figure}[t]
	\centering
	\subfigure{
		\includegraphics[scale=0.4]{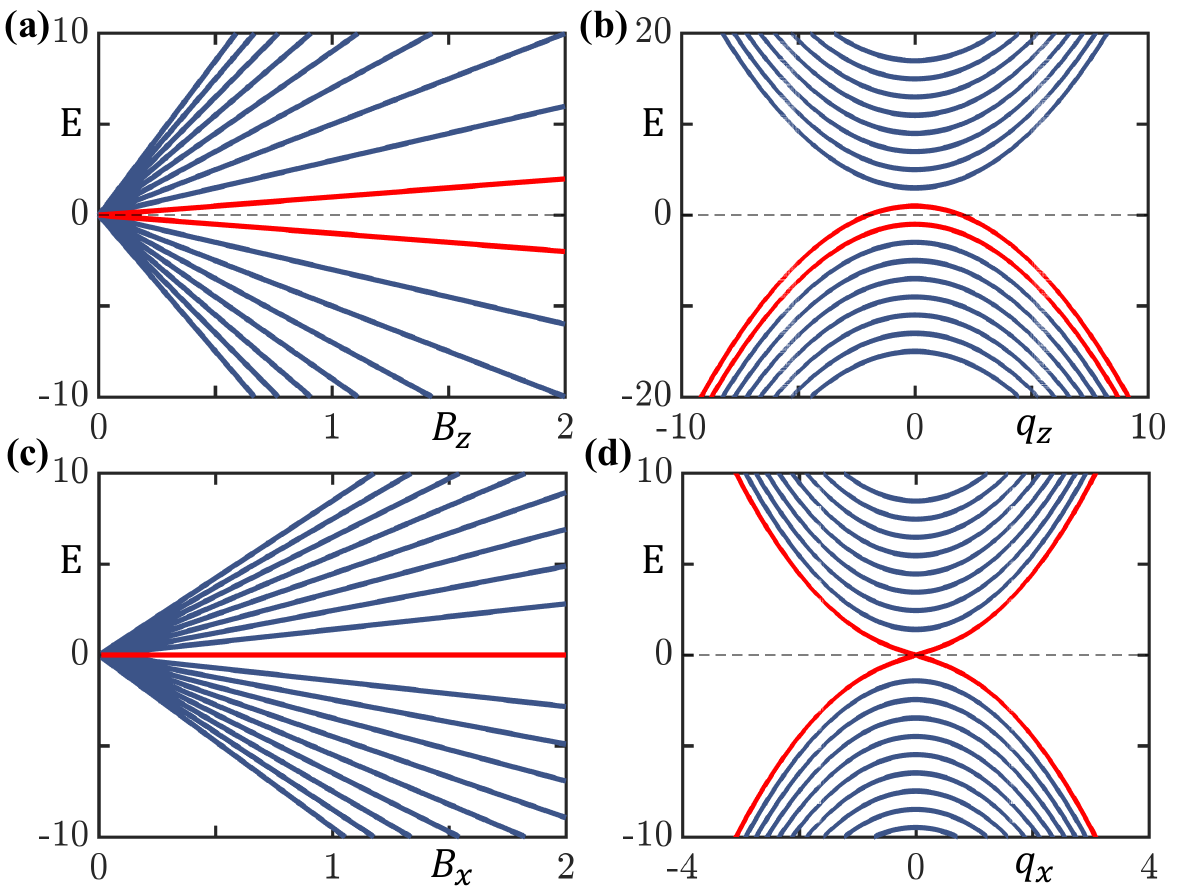}}
	\caption{(a-b) Landau levels under a $z$-direction magnetic field. (c-d) Landau levels under an $x$-direction magnetic field.
 (a) $q_{z}=0$, (b) $B_{z}=1$, (c) $q_{x}=0$, (d) $B_{x}=1$. Shared parameters are $v=1$ and $v_{z}=0.5$.  $e=\hbar=1$ is adopted
 in all figures.}
	\label{Fig4}
\end{figure}

{\it Anisotropic behaviors of Landau levels.---}Landau levels (LLs) are of particular interest in
the study of topological semimetals as they can reveal the Berry phase
of the Fermi surface~\cite{Mikitik1999Berry,Zhang2005graphene,Novoselov2006}.
Since the dipole Berry curvature is
quite anisotropic,
one can expect that remarkable anisotropic behaviors will show up in
the LL structure, even when the zero-field energy spectra are isotropic.
Indeed, based on the low-energy Berry-dipole Hamiltonian, we find
that, when the magnetic field is along the rotation symmetry axis, i.e., $z$ axis,
the LLs are analytically solvable and the result is~\cite{supplemental}
\begin{eqnarray}
	E_{\chi,0}(q_{z})&=&v^{2}/l_{B}^{2}-v_{z}^{2}q_{z}^{2},\nonumber\\
E_{\chi,n\pm}(q_{z})&=&v^{2}/l_{B}^{2}\pm(2nv^{2}/l_{B}^{2}+v_{z}^{2}q_{z}^{2}),\, n\geq1,
\end{eqnarray}
where $l_{B}=\sqrt{\hbar/eB}$ is the magnetic length and $n$ is an integer labeling the LLs.
A salient feature of the above LLs is that all of them are separated and have a linear dependence on the field strength,
as shown in Figs.\ref{Fig4}(a) and \ref{Fig4}(b). In contrast, when the magnetic field changes to
the $x$ direction, we find  that the exact LLs cannot be analytically obtained, but
in the small $q_{x}$ regime, the LLs take the approximate form~\cite{supplemental}
\begin{eqnarray}
E_{\chi,1\pm}(q_{x})&=&\pm\sqrt{2v^{3}v_{z}/l_{B}^{2}}q_{x},\nonumber\\
E_{\chi,n\pm}(q_{x})&=&\pm[F_{n}vv_{z}/l_{B}^{2}+G_{n}v^{2}q_{x}^{2}], \, n\geq2,
\end{eqnarray}
where $F_{n}=2\sqrt{n(n-1)}$, $G_{n}=(2n-1)/2\sqrt{n(n-1)}$. Compared to the
$z$-direction case, a remarkable difference is that the two middle
LLs cross at $q_{x}=0$. The numerically determined
exact LLs show that the two middle LLs form a 1D nonlinear
Dirac cone, as shown in Figs.\ref{Fig4}(c) and \ref{Fig4}(d).

{\it Discussions and conclusions.---}We have introduced BDSMs into the
family of topological semimetals and shown that they carry many salient
characteristics both in boundary physics and in bulk physics. This concept can also
be generalized to superconducting systems just like other semimetals~\cite{Yang2014dirac}.
Although condensed-matter material realizations might be challenging,
the implementation of the proposed BDSMs is realistic in many artificial systems with
high flexibility, such as circuit systems which have recently realized the very related
Hopf insulators~\cite{Wang2023hopf}.

{\it Acknowledgements.---}
We would like to thank Prof. Qian Niu and Prof. Zhi Wang for helpful discussions.
This work is supported by the National Natural Science Foundation of China (Grant No.12174455),
the Natural Science Foundation of Guangdong Province
(Grant No. 2021B1515020026), and the Guangdong
Basic and Applied Basic Research Foundation (Grant No.
2023B1515040023).

\bibliography{dirac}

\begin{widetext}
\clearpage
\begin{center}
\textbf{\large Supplemental Material for ``Berry-dipole Semimetals''}\\
\vspace{4mm}
{Zheng-Yang Zhuang, Chaoyi Zhang, Xiao-Jiao Wang, Zhongbo Yan$^{*}$}\\
\vspace{2mm}
{\em Guangdong Provincial Key Laboratory of Magnetoelectric Physics and Devices, \\State Key Laboratory of Optoelectronic Materials and Technologies,\\School of Physics, Sun Yat-sen University, Guangzhou 510275, China}\\
\end{center}

\setcounter{equation}{0}
\setcounter{figure}{0}
\setcounter{table}{0}
\makeatletter
\renewcommand{\theequation}{S\arabic{equation}}
\renewcommand{\thefigure}{S\arabic{figure}}
\renewcommand{\bibnumfmt}[1]{[S#1]}

The supplemental material contains detailed calculations for various physical properties near the Berry-dipole node, including the anomalous Hall conductivity, orbital magnetic moment, orbital magnetization, and the Landau levels. Three sections are in order: (I) Hamiltonian and band topology; (II) Topological and physical properties associated with the low-energy Berry-dipole Hamiltonains; (III) Anisotropic Landau levels for the low-energy Berry-dipole Hamiltonians.

\section{I. Hamiltonian and band topology}

In the main text, the two-band Hamiltonian for the 3D Berry-dipole semimetal is constructed by the Hopf map
\begin{eqnarray}
	\mathcal{H}(\bk)=\bd(\bk)\cdot\boldsymbol{\sigma},
\end{eqnarray}
where $\boldsymbol{\sigma}=(\sigma_{x},\sigma_{y},\sigma_{z})$ are the Pauli matrices, and $\bd=(d_{x},d_{y},d_{z})$ with each component determined by $d_{i}=\zeta^\dagger\sigma_{i}\zeta$.  $\zeta$ is a complex spinor, i.e., $\zeta=(\zeta_{1},\zeta_{2})^{\rm T}$. In this work we choose
\begin{eqnarray}
	\zeta_{1}&=&\lambda(\sin k_{x}+i\sin k_{y}),\\
	\zeta_{2}&=&\delta(\cos k_{x}-\cos k_{y})+i(M-t\sum_{i=x,y,z}\cos k_{i}).
\end{eqnarray}
The explicit form of the resultant Hamiltonian reads
\begin{eqnarray}
	\mathcal{H}(\bk)&=&2\lambda\left[\delta\sin k_{x}(\cos k_{x}-\cos k_{y})+\sin k_{y} (M-t\sum_{i=x,y,z}\cos k_{i})\right]\sigma_{x}\nonumber\\
	&&-2\lambda\left[\delta\sin k_{y}(\cos k_{x}-\cos k_{y})-\sin k_{x}(M-t\sum_{i=x,y,z}\cos k_{i})\right]\sigma_{y}\nonumber\\
	&&+\left[\lambda^{2}(\sin^{2}k_{x}+\sin^{2}k_{y})-\delta^{2}(\cos k_{x}-\cos k_{y})^{2}-(M-t\sum_{i=x,y,z}\cos k_{i})^{2}\right]\sigma_{z},
	\label{eq: latticeH}
\end{eqnarray}
and the corresponding energy spectra are
\begin{eqnarray}
	E_{\pm}(\bk)=\pm\left[\lambda^{2}(\sin^{2}k_{x}+\sin^{2} k_{y})+\delta^{2}(\cos k_{x}-\cos k_{y})^{2}+(M-t\sum_{i=x,y,z}\cos k_{i})^{2}\right].
\end{eqnarray}

Let us first consider the case with $\delta=0$. Accordingly, the Hamiltonian reduces to
\begin{eqnarray}
	\mathcal{H}(\bk)&=&2\lambda\sin k_{y} (M-t\sum_{i=x,y,z}\cos k_{i})\sigma_{x}
	+2\lambda\sin k_{x}(M-t\sum_{i=x,y,z}\cos k_{i})\sigma_{y}\nonumber\\
	&&+\left[\lambda^{2}(\sin^{2}k_{x}+\sin^{2}k_{y})-(M-t\sum_{i=x,y,z}\cos k_{i})^{2}\right]\sigma_{z}.
	\label{eq: H_delta0}
\end{eqnarray}
It is easy to find that the above form has $C_{4z}$ rotation symmetry. This discrete symmetry will extend to a continuous
$U(1)$ rotation symmetry in the continuum limit (consider the low-energy counterpart of the lattice Hamiltonian), which
suggests that the boundary physics will not depend on the direction with open boundary conditions. Focusing on the bulk, it can readily be
 found that, if $|M/3t|\leq1$,  doubly degenerate band nodes exist and they are located at one or two of the four high-symmetry $k_{z}$ lines passing $(k_{x},k_{y})=(0/\pi,0/\pi)$. Without loss of generality, we consider $M\in (t,3t)$ with $t$ chosen to be positive. For this case, two band nodes are located at $\bk_{n,\chi}=\chi(0,0,k_{0})$, where $k_{0}=\arccos(M/t-2)$, and $\chi=\pm1$ refers to the two nodes respectively.

A convenient way to understand the topological property of the 3D Hamiltonian in Eq.(\ref{eq: H_delta0}) is to view it as
a stacking of 2D Hamiltonian along the $k_{z}$ direction. In this way, $k_{z}$ acts as a parameter of a 2D Hamiltonian
$\mathcal{H}(k_{x},k_{y})$. For the convenience of discussion, we introduce the notation
$\mathcal{H}_{k_{z}}(k_{x},k_{y})$ to denote the 2D Hamiltonian at a given $k_{z}$.
For $\mathcal{H}_{k_{z}}(k_{x},k_{y})$,  it belongs to the A class in the ten-fold way classification~\cite{Schnyder2008,kitaev2009periodic}, thereby it is characterized by the first-class Chern number as long as it is fully gapped.  In terms
of the formula,
\begin{eqnarray}
\mathcal{C}_{1}(k_{z})=\frac{1}{4\pi}\int_{-\pi}^{\pi}dk_{x}\int_{-\pi}^{\pi}dk_{y}
\frac{\bd(\bk)\cdot[\partial_{k_{x}}\bd(\bk)\times\partial_{k_{y}}\bd(\bk)]}{d^{3}(\bk)},
\end{eqnarray}
where $d(\bk)=|\bd(\bk)|$, one can find that the Chern number $\mathcal{C}_{1}(k_{z})$ is always zero for these gapped
planes, i.e., $k_{z}\neq \pm k_{0}$. The vanishing of Chern number suggests that the 2D Hamiltonian at a given
$k_{z}$ does not host chiral edge states. However, we find that this is not equivalent to stating that
$\mathcal{H}_{k_{z}}(k_{x},k_{y})$ is topologically trivial.

We find that there are two ways to see
that the system in fact carries nontrivial topology. The first way is based on a dimension reduction perspective.
Concretely, we note that the 2D Hamiltonian
has chiral symmetry along the high symmetry lines with $k_{x}=0/\pi$ or  $k_{y}=0/\pi$, so that a winding number can be assigned
to characterize the effective 1D Hamiltonian $\mathcal{H}_{k_{z}}(k_{x}=0/\pi,k_{y})$ or $\mathcal{H}_{k_{z}}(k_{x},k_{y}=0/\pi)$.
In terms of the formula\cite{Ryu2010},
\begin{eqnarray}
W^{(x)}_{0/\pi}(k_{z})&=&\frac{1}{4\pi i}\int_{-\pi}^{\pi}dk_{y}\text{Tr}[\mathcal{S}_{x}\mathcal{H}_{k_{z}}^{-1}(k_{x}=0/\pi,k_{y})
\partial_{k_{y}}\mathcal{H}_{k_{z}}(k_{x}=0/\pi,k_{y})],\nonumber\\
W^{(y)}_{0/\pi}(k_{z})&=&\frac{1}{4\pi i}\int_{-\pi}^{\pi}dk_{x}\text{Tr}[\mathcal{S}_{y}\mathcal{H}_{k_{z}}^{-1}(k_{x},k_{y}=0/\pi)
\partial_{k_{x}}\mathcal{H}_{k_{z}}(k_{x},k_{y}=0/\pi)],
\end{eqnarray}
where $\mathcal{S}_{x}=\sigma_{y}$ satisfies $\{\mathcal{S}_{x}, \mathcal{H}_{k_{z}}(k_{x}=0/\pi,k_{y})\}=0$
and $\mathcal{S}_{y}=\sigma_{x}$ satisfies $\{\mathcal{S}_{y}, \mathcal{H}_{k_{z}}(k_{x},k_{y}=0/\pi)\}=0$, one obtains
\begin{eqnarray}
&|W^{(x)}_{0}(k_{z})|=|W^{(y)}_{0}(k_{z})|=2, \quad -k_{0}<k_{z}<k_{0}; \nonumber\\
&W^{(x)}_{0}(k_{z})=W^{(y)}_{0}(k_{z})=0, \quad k_{0}<|k_{z}|\leq\pi; \nonumber\\
&W^{(x)}_{\pi}(k_{z})=W^{(y)}_{\pi}(k_{z})=0, \quad -\pi <k_{z}\leq\pi.
\end{eqnarray}
The nontrivial values of $W^{(x)}_{0}(k_{z})$ and $W^{(y)}_{0}(k_{z})$ for $-k_{0}<k_{z}<k_{0}$ indicate
that there are two bound states at one $x(y)$-normal edge when $k_{y}(k_{x})=0$.
Combining the fact that the Chern number is always trivial, one can immediately figure out
that there should exist a pair of counter-propagating edge states, which can simultaneously
be consistent with the absence of chiral edge states and the existence of two bound states
at a given boundary momentum. By calculating the energy spectra under open boundary conditions
in the $x$ direction for a fixed $k_{z}$ in the regime $-k_{0}<k_{z}<k_{0}$, we do
find the expected pair of counter-propagating edge states, as shown in Fig.\ref{Figs1}(a). In contrast,
when $k_{0}<|k_{z}|\leq\pi$, no edge state is found (see Fig.\ref{Figs1}(b)), which is consistent with the winding number
obtained. The isoenergy contours of the pair of counter-propagating edge states form the two surface Fermi arcs
connecting the projections of the bulk Fermi surfaces enclosing the Berry-dipole nodes.

\begin{figure}[t]
	\centering
	\subfigure{
		\includegraphics[scale=0.6]{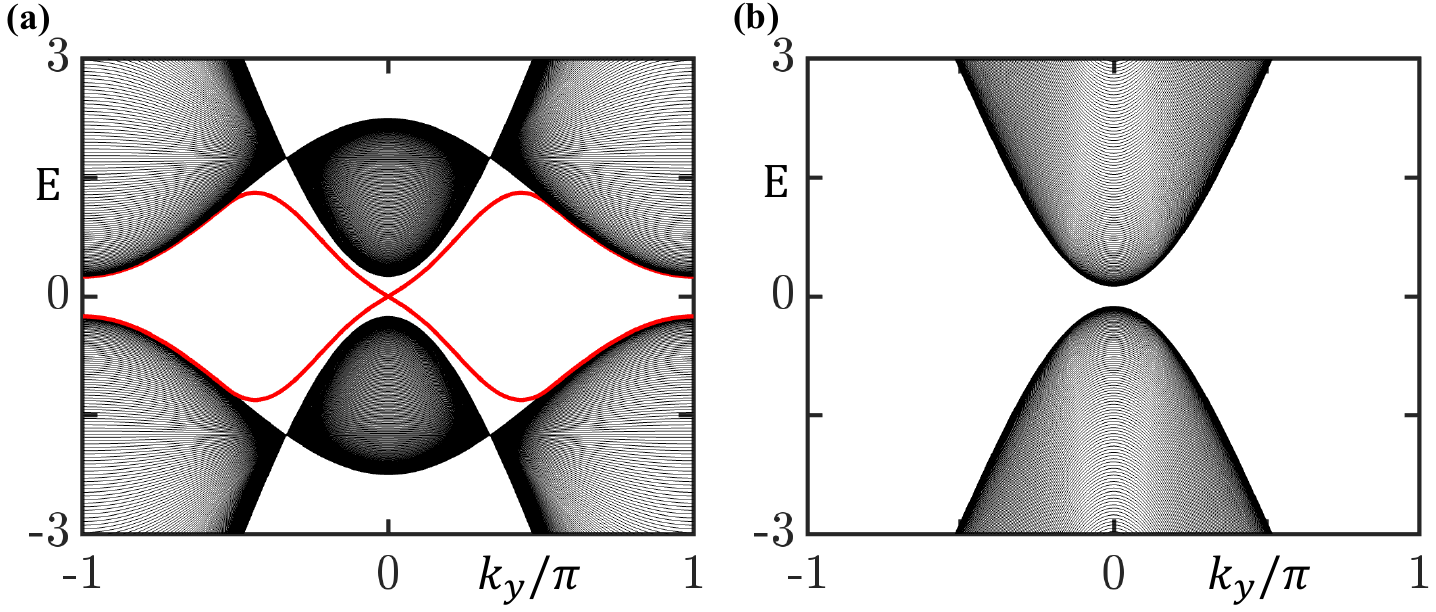}}
	\caption{Energy spectra at a given $k_{z}$. Open boundary conditions are applied in the $x$ direction. (a) $k_{z}=0$, a pair of counter-propagating gapless edge states is found. (b) $k_{z}=k_{0}+0.5$, there is no edge state. Other parameters are $M=1.5$, $t=\lambda=1$ and $\delta=0$. For this set of parameters, $k_{0}=\frac{2 \pi}{3}$.}
	\label{Figs1}
\end{figure}

\begin{figure}[t]
	\centering
	\subfigure{
		\includegraphics[scale=0.6]{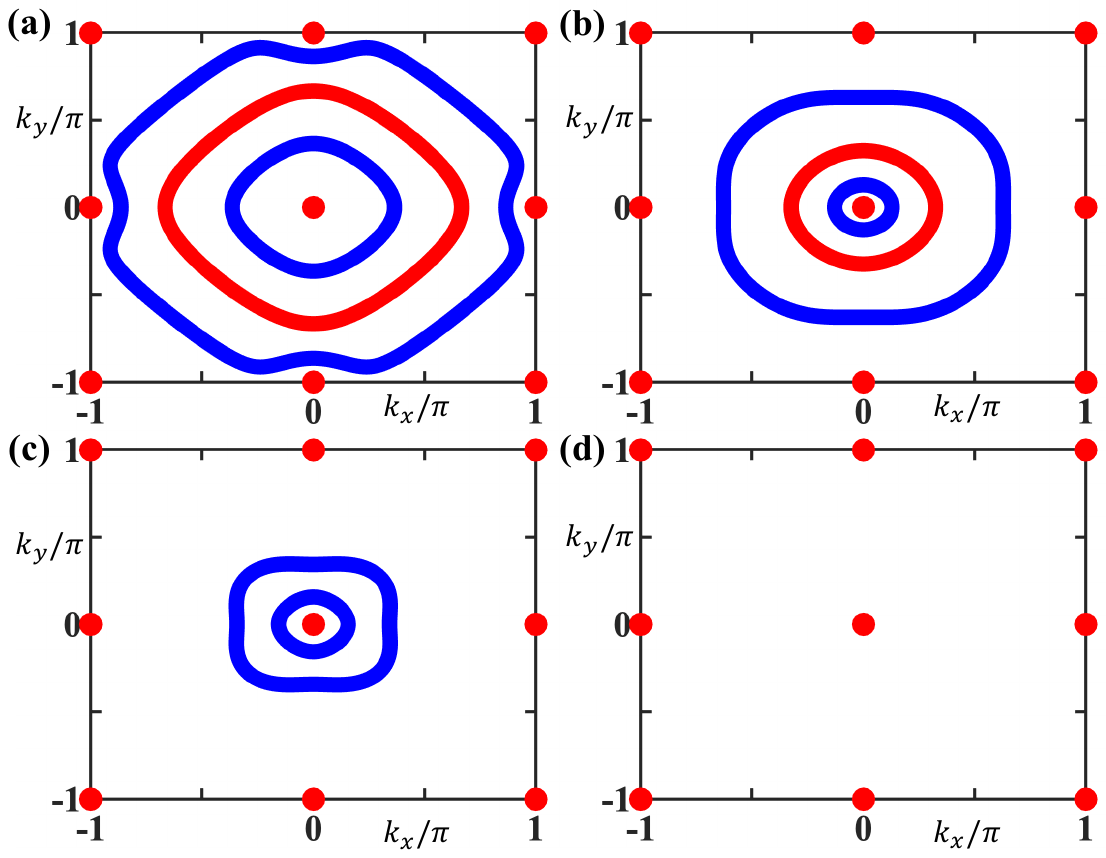}}
	\caption{Zero-value contours of the $d_{x}$, $d_{y}$ and $d_{z}$ terms at a given $k_{z}$ plane. The blue lines
refer to the contours satisfying $d_{z}=0$, and the red lines and dots refer to the contours satisfying $d_{x}=d_{y}=0$. The two configurations in (a) and (b) are topological since the blue lines cannot be annihilated unless crossing the red line when $|k_{z}|<k_{0}$. In contrast, when $k_{0}<|k_{z}|\leq\pi$, the annihilation of the two blue lines is feasible, as shown in (c). In (d), the zero-energy contours for $d_{z}=0$
are already annihilated. In (a-d), the values of $k_{z}$ are $0$, $k_{0}-0.5$, $k_{0}+0.5$, and $\pi$ in order. Other parameters are the same as those in Figs.\ref{Figs1}.}\label{Figs2}
\end{figure}

The second way to see that the 2D Hamiltonian $\mathcal{H}_{k_{z}}(k_{x},k_{y})$ has nontrivial topology is through
the configurations of the zero-value contours of the terms in the Hamiltonian. The logic behind is that all terms of the Hamiltonian
are anticommuting with each other, so the bulk energy gap can get closed only when their zero-value contours share
common points. On the other hand, it is known that the change of first-order topology must undergo a closure of the bulk energy gap.
Therefore, if the zero-value contours of the $d_{x,y,z}$ terms form a configuration that cannot be adiabatically (without closing
the bulk energy gap, or equivalently, the zero-value contours do not cross at a common point) deformed to a trivial configuration,
the 2D Hamiltonian describes a topological insulator phase. What is a trivial configuration?
It is known that a trivial phase means that it is adiabatically connected to an atomic insulator. The simplest atomic insulator corresponds
to the absence of hopping, and only on-site potentials or couplings within one unit cell exist. For such an atomic insulator,
if it is described by a two-band Hamiltonian, no $d_{x,y,z}$ terms will have zero-value contours.  Therefore, a trivial configuration
means that the zero-value contours of the $d_{x,y,z}$ terms can be adiabatically deformed to vanish without the closure of energy gap.

For the two-band Hamiltonians concerned here, to see whether the configuration is topological or trivial, we find it is most convenient to plot the zero-value contours satisfying $d_{z}=0$ and the ones simultaneously satisfying $d_{x}=d_{y}=0$. The former in general is lines
in the 2D Brillouin zone, and the latter is in general points. Representative configurations for $-k_{0}<k_{z}<k_{0}$ are
shown in Figs.\ref{Figs2}(a) and \ref{Figs2}(b), and representative configurations for $k_{0}<|k_{z}|\leq\pi$ are shown in Figs.\ref{Figs2}(c) and \ref{Figs2}(d).
From the two figures in the upper row, it is easy to see that if one fixes the zero-value
contours satisfying $d_{x}=d_{y}=0$, the two rings satisfying $d_{z}=0$ cannot be deformed to vanish without crossing
the contours satisfying $d_{x}=d_{y}=0$. The crossing, however, corresponds to a closure of the bulk energy gap. This indicates
that the configurations in Figs.\ref{Figs2}(a) and \ref{Figs2}(b) are conditionally obstructed to a trivial configuration.
Therefore, the Hamiltonian
corresponding to configurations  in Figs.\ref{Figs2}(a) and \ref{Figs2}(b) must  be topologically nontrivial in some way. As revealed by numerical calculations,
the nontrivial topology manifests as the existence of a pair of counter-propagating gapless edge states even though the
system does not have time-reversal symmetry.  In contrast, the configurations
in Figs.\ref{Figs2}(c) and \ref{Figs2}(d) apparently belong to the class of trivial configurations, and numerical calculations also confirm
the absence of any type of gapless edge states.

Now we discuss the effects of  the terms involving $\delta$. When $\delta$ becomes nonzero, the chiral symmetry on the high symmetry lines $k_{x}=0/\pi$
or $k_{y}=0/\pi$ are broken. As a consequence of the lift of the protecting symmetry, the pair of counter-propagating gapless edge states are gapped, as shown in Fig.\ref{Figs3}(a). However, the system remains to be topologically nontrivial. In the regime $-k_{0}<k_{z}<k_{0}$, we find
that the $\mathcal{H}_{k_{z}}(k_{x},k_{y})$ now describes a second-order topological insulator~\cite{Yan2019hosca}. When both $x$ and $y$ directions
take open boundary conditions, there are four zero-energy states, one per corner, as shown in Fig.\ref{Figs3}(b). Again plotting the
configurations of the zero-value contours of the $d_{x,y,z}$ terms (see Fig.\ref{Figs3}(c)), we find that, similar to the case with $\delta=0$,
the configuration remain to be conditionally obstructed to a trivial configuration. The only difference between
the two cases with $\delta=0$ and $\delta\neq0$ is that for the former, the annihilation of the two rings
satisfying $d_{z}=0$ are obstructed by a 1D ring satisfying $d_{x}=d_{y}=0$, while for the latter,
their annihilation is obstructed by four 0D points satisfying $d_{x}=d_{y}=0$. Interestingly,
the dimension of the zero-value contours lying between the two zero-value contours satisfying $d_{z}=0$
has a one-to-one correspondence with the dimension of the boundary states. The result indicates that once the
$\delta$ term is finite, the system transits from a first-order BDSM to a second-order BDSM.
Since all these 2D planes for $-k_{0}<k_{z}<k_{0}$ correspond to a second-order topological insulator with corner states,
these corner states consist of four hinge arcs at the charge neutrality condition after reconsidering the system as a 3D system.

\begin{figure}[t]
	\centering
	\subfigure{
		\includegraphics[scale=0.6]{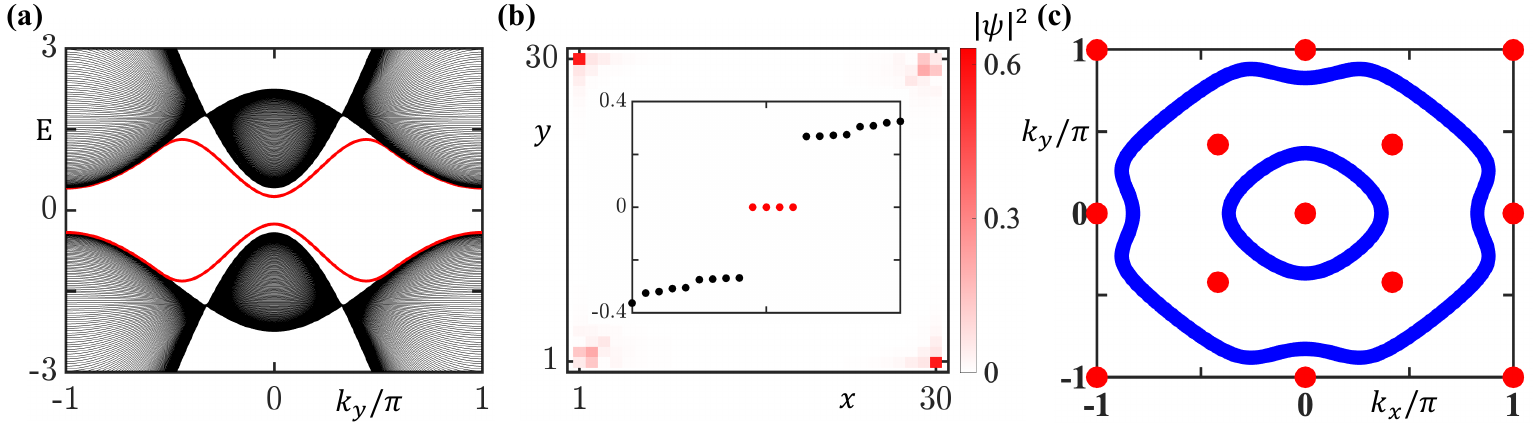}}
	\caption{(a) Energy spectra with nonzero $\delta$ and open (periodic) boundary conditions in the $x$ ($y$) direction. The edge states become gapped. (b) Four zero-energy states (red dots in the inset) are found when open boundary conditions are set in both $x$ and $y$ directions. The probability density profile of the four zero-energy states are sharply localized at the corners.
(c) The zero-value contours of $d_{x}$, $d_{y}$ and $d_{z}$. The four red dots between the two blue lines obstruct their annihilation without
the closure of energy gap. The parameters in (a-c) are $M=1.5$, $t=\lambda=1$, $k_{z}=0$, and $\delta=0.2$.}\label{Figs3}
\end{figure}

\section{II. Topological and physical properties associated with the low-energy Berry-dipole Hamiltonains}

\subsection{A. Topological property of the Berry-dipole nodes}

Above we have shown that when $\delta\neq0$, the topological boundary states have a dramatic change in comparison with
the case with $\delta=0$. However,
when focusing on the low-energy physics near the band nodes, the terms involving $\delta$ are unimportant since their leading-order contribution to the energy spectra is  of fourth order in momentum, while the leading-order contributions from other terms are second order in momentum. Therefore,  we disregard all terms
involving $\delta$.  Performing a Taylor expansion near the band nodes at $\bk_{n,\chi}$ up to second order in momentum, one obtains
\begin{eqnarray} \mathcal{H}_{\chi}(\bq)=\chi2vv_{z}q_{y}q_{z}\sigma_{x}+\chi2vv_{z}q_{x}q_{z}\sigma_{y}+\left[v^{2}(q_{x}^{2}+q_{y}^{2})-v_{z}^{2}q_{z}^{2}\right]\sigma_{z},
	\label{eq: lowH}
\end{eqnarray}
where $v=\lambda$, $v_{z}=t\sqrt{1-(M/t-2)^2}$, $\bq=(q_{x},q_{y},q_{z})$ denotes the momentum measured from
the corresponding band node, and $\chi=\pm$ refer to the two band nodes located at $\bk_{n,\chi}=\chi(0,0,\arccos(M/t-2))$. The energy spectra of
the low-energy continuum Hamiltonian are
\begin{eqnarray}
	E_{\chi,\pm}(\bq)=\pm\left[v^{2}(q_{x}^{2}+q_{y}^{2})+v_{z}^{2}q_{z}^{2}\right],
\end{eqnarray}
where $\pm$ refer to the conduction and valence bands, respectively. Obviously the energy spectra are quadratic in all directions. Accordingly, the Fermi velocities near the nodes are linear in momentum,
\begin{eqnarray}
	\bv_{\chi,\pm}(\bq)=\frac{\partial E_{\chi,\pm}(\bq)}{\partial \bq}=\pm2(v^{2}q_{x},v^{2}q_{y},v_{z}^{2}q_{z}).
\end{eqnarray}
Noteworthily,  $\bv_{\chi,\pm}(\bq)$ is independent of $\chi$ as the energy spectra do not depend on $\chi$.

Despite the fact that the two low-energy Hamiltonians associated with the two nodes are different, the Berry curvatures are the same.
In 3D, the Berry curvature has three components and can be viewed as a vector. The Berry curvature vector near the nodes is
\begin{eqnarray}
	\bOmega_{\chi}^{(c)}(\bq)=-\bOmega_{\chi}^{(v)}(\bq)=-\frac{2v^{2}v_{z}^{2}q_{z}}{E_{\chi,+}^{2}}\bq,
\end{eqnarray}
where the superscript ``c(v)'' refers to the conduction (valence) band. Interestingly,
the momentum dependence of the Berry curvature resembles the field distribution of a dipole.
Since the Berry curvatures for the conduction and valence bands are exactly opposite for the two-band model, below
we focus on the conduction band for a detailed investigation.  It is easy to find that an integral of the Berry curvature
over an isoenergy close surface identically vanishes, i.e.,
\begin{eqnarray}
	C&=&\frac{1}{2\pi}\oiint_{E_{\chi,+}=\mu}\bOmega_{\chi}^{(c)}\cdot d\bS\nonumber\\
&=&-\frac{1}{2\pi}\oiint_{E_{\chi,+}=\mu}\frac{2v^{2}v_{z}^{2}q_{z}}{[v^{2}(q_{x}^{2}+q_{y}^{2})+v_{z}^{2}q_{z}^{2}]^{2}}\bq\cdot d\bS\nonumber\\
&=&-\frac{1}{2\pi\mu^{2}}\oiint_{E_{\chi,+}=\mu}2v^{2}v_{z}^{2}q_{z}\bq\cdot d\bS,\nonumber\\
&=&-\frac{1}{2\pi\mu^{2}}\iiint_{E_{\chi,+}<\mu}\nabla\cdot(2v^{2}v_{z}^{2}q_{z}\bq)dq_{x}dq_{y}dq_{z}\nonumber\\
&=&-\frac{1}{2\pi\mu^{2}}\iiint_{E_{\chi,+}<\mu}8v^{2}v_{z}^{2}q_{z}dq_{x}dq_{y}dq_{z}\nonumber\\
&=&-\frac{8}{2\pi\mu^{2}}\iiint_{E_{\chi,+}<\mu}\tilde{q}_{z}d\tilde{q}_{x}d\tilde{q}_{y}d\tilde{q}_{z}.
	\label{def: Hopf point1}
\end{eqnarray}
At the last step, we have rescaled the momentum, i.e., $\tilde{q}_{x}=vq_{x}$, $\tilde{q}_{y}=vq_{y}$,
and $\tilde{q}_{z}=v_{z}q_{z}$. Accordingly, $E_{\chi,+}=v^{2}(q_{x}^{2}+q_{y}^{2})+v_{z}^{2}q_{z}^{2}=\tilde{q}_{x}^{2}
+\tilde{q}_{y}^{2}+\tilde{q}_{z}^{2}$, suggesting that the volume is enclosed by a sphere in the rescaled momentum
space.  Changing the Cartesian coordinates into the spherical coordinates, one obtains the integral
\begin{eqnarray}
C&=&-\frac{8}{2\pi\mu^{2}}\iiint_{E_{\chi,+}<\mu}\tilde{q}_{z}d\tilde{q}_{x}d\tilde{q}_{y}d\tilde{q}_{z}\nonumber\\
&=&-\frac{8}{2\pi\mu^{2}}\int_{0}^{\sqrt{\mu}}d\tilde{q}\int_{0}^{\pi}d\theta\int_{0}^{2\pi}d\phi \tilde{q}^{3}\cos\theta\sin\theta\nonumber\\
&=&-\int_{0}^{\pi}d\theta\sin2\theta=0.
\end{eqnarray}
The vanishment of $C$ suggests that the net Berry monopole charge within the surface is zero. However,
it is easy to see that the integral over the upper-half surface ($q_{z}\geq0$, or say $0\leq\theta\leq\pi/2$) or
the lower-half surface ($q_{z}\leq0$, or say $\pi/2\leq\theta\leq\pi$) is quantized to a nonzero value,
\begin{eqnarray}
	C_{+}&=&-\int_{0}^{\pi/2}d\theta\sin2\theta=-1,\nonumber\\
	C_{-}&=&-\int_{\pi/2}^{\pi}d\theta\sin2\theta=1.
	\label{def: Hopf point2}
\end{eqnarray}
The result suggests that the band node acts as a point dipole which, pictorially, is formed
by the overlap of one monopole and one antimonopole in a mirror symmetric way~\cite{Nelson2021hopf}.

\subsection{B. Intrinsic anomalous Hall effect}

As the Berry-dipole nodes carry nontrivial Berry curvatures, a direct physical observable effect is the intrinsic anomalous
Hall effect. In 3D, the general formula for the Hall current contributed by Berry curvature is given by
\begin{eqnarray}
	j_{a}&=&\frac{e^{2}}{\hbar}\sum_{n}\iiint\frac{d^{3}q}{(2\pi)^{3}}f(E_{n}(\bq))\varepsilon_{abc}\Omega_{b}^{(n)}\mathcal{E}_{c},
	\label{def: anomalous current}
\end{eqnarray}
where the subscript $a$ of $j_{a}$ labels the direction of the current, $n$  the band index,
$f$ the Fermi-Dirac distribution function, and  $\mathcal{E}_{c}$ the component of the electric
field in the $c$ direction. Here we focus on the zero-temperature limit, so that $f$ is reduced
as a step function, i.e., $f(E_{n}(\bq))=\Theta(\mu-E_{n}(\bq))$. Again viewing the Hamiltonian
as a parameter dependent 2D Hamiltonian, it is easy to find that the 2D
Hamiltonians $\mathcal{H}_{k_{x}}(k_{y},k_{z})$ and $\mathcal{H}_{k_{y}}(k_{x},k_{z})$ have
an effective spinless time-reversal symmetry.
This suggests that the Hall conductivities  $\sigma_{yz}$ and $\sigma_{xz}$ vanish identically. Only
the 2D Hamiltonians $\mathcal{H}_{k_{z}}(k_{x},k_{y})$ lack time-reversal symmetry,
so only $\sigma_{xy}$ can be nonzero. Also focusing on a positive and small
$\mu$, since all gapped $k_{z}$ planes have trivial Chern number, the occupied band
has vanishing contribution to $\sigma_{xy}$. Therefore, we can just focus on
the upper band. In the weakly-doped regime, $\sigma_{xy}$ can be determined by
using the low-energy Hamiltonians around the two band nodes. Since the two
band nodes have the same Berry curvature and energy spectra, they have the same
contribution to $\sigma_{xy}$. The result is thereby given by
\begin{eqnarray}
\sigma_{xy}(\mu)&=&\sum_{\chi}\frac{e^{2}}{\hbar}\iiint \frac{dq_{x}dq_{y}dq_{z}}{(2\pi)^{3}}\Omega_{\chi}^{(c)}(\bq)\Theta(\mu-E_{\chi,+}(\bq))\nonumber\\
&=&\frac{2e^{2}}{\hbar}\iiint \frac{dq_{x}dq_{y}dq_{z}}{(2\pi)^{3}}\Omega_{+}^{(c)}(\bq)\Theta(\mu-E_{+,+}(\bq))\nonumber\\
&=&-\frac{2e^{2}}{\hbar}\iiint \frac{dq_{x}dq_{y}dq_{z}}{(2\pi)^{3}}\frac{2v^{2}v_{z}^{2}q_{z}^{2}}{[v^{2}(q_{x}^{2}+q_{y}^{2})
+v_{z}^{2}q_{z}^{2}]^{2}}\Theta(\mu-E_{+,+}(\bq))\nonumber\\
&=&-\frac{2e^{2}}{\hbar v_{z}}\iiint \frac{d\tilde{q}_{x}d\tilde{q}_{y}d\tilde{q}_{z}}{(2\pi)^{3}}\frac{2\tilde{q}_{z}^{2}}{\tilde{q}^{4}}\Theta(\mu-\tilde{q}^{2})\nonumber\\
&=&-\frac{2e^{2}}{(2\pi)^{3}\hbar v_{z}}\int_{0}^{\sqrt{\mu}}d\tilde{q}\int_{0}^{\pi}d\theta\int_{0}^{2\pi}d\phi \frac{2\tilde{q}^{4}\cos^{2}\theta\sin\theta}{\tilde{q}^{4}}\nonumber\\
&=&-\frac{2e^{2}}{(2\pi)^{3}\hbar v_{z}}\frac{8\pi\sqrt{\mu}}{3}=-\frac{e^{2}}{h}\frac{4\sqrt{\mu}}{3\pi v_{z}}.\label{Hall}
\end{eqnarray}
Interestingly, $\sqrt{\mu}/v_{z}$ has the geometric interpretation as the radius of the Fermi surface in the $z$ direction. Defining
the diameter parameter $D_{F,z}\equiv2\sqrt{\mu}/v_{z}$, we reach the expression in Eq.(11) of the main text,
\begin{eqnarray}
\sigma_{xy}(\mu)=-\frac{e^{2}}{h}\frac{2D_{F,z}}{3\pi}.
\end{eqnarray}
Before ending this section, it might be interesting to give a comparison with the Hall conductivity
in an ideal Weyl semimetal with only one pair of Weyl nodes at the same energy. For the latter, the Hall
conductivity also has a connection with one geometric quantity, the distance between the two Weyl nodes~\cite{yang2011}.
Concretely,  when the Fermi energy crosses the Weyl nodes, it is known that the Hall conductivity is proportional
to the separation between the two Weyl nodes in the momentum space, i.e.,
\begin{eqnarray}
\sigma_{ij}=\frac{e^{2}}{h}\epsilon_{ijl}\frac{\Delta k_{l}}{2\pi}
\end{eqnarray}
where $\Delta k_{l}$ denotes the separation between the two Weyl nodes in
the $l$ direction. If one varies the chemical potential but still keeps the system in the weakly
doped regime, $\sigma_{ij}$ will retain its value since the low-energy Weyl Hamiltonian of the form,
$\mathcal{H}(\bq)=\sum_{ij}v_{ij}q_{i}\sigma_{j}$,
has emergent time-reversal symmetry which forces the vanishment of the leading-order contributions from
the Berry curvature.

\subsection{C. Orbital magnetic moment and orbital magnetization}

In the semiclassical theory, the self-rotation of a wave packet leads to an orbital magnetic moment. The orbital magnetic moment
is given by~\cite{Xiao2010BP}
\begin{eqnarray}
	\boldsymbol{m}(\bq)=-i\frac{e}{2h}\bra{\nabla_{\bq}u}\times\left[\mathcal{H}(\bq)-E(\bq)\right]\ket{\nabla_{\bq}u},
\end{eqnarray}
where $\ket{u(\bq)}$ and $E(\bq)$ are the cell-periodic Bloch function and eigenenergy of the Hamiltonian $\mathcal{H}(\bq)$,
respectively. For a two-band Hamiltonian, $\mathcal{H}(\bq)=\bd(\bq)\cdot\bsigma$, $\boldsymbol{m}(\bq)$ has a simple relation
with the Berry curvature~\cite{Xiao2007MM},
\begin{eqnarray}
	m_{\alpha}(\bq)=\frac{ed(\bq)}{\hbar}\Omega_{\alpha}(\bq),\label{OMMBC}
\end{eqnarray}
where $d(\bq)=|\bd(\bq)|$, and $\Omega_{\alpha}(\bq)$ is the $\alpha$ component of the Berry curvature vector.
Let us also focus on the low-energy Berry-dipole Hamiltonian and the upper band. Since the Berry curvatures
and energy spectra for the two low-energy Berry-dipole Hamiltonians are also the same,  their orbital magnetic moments are
the same. Focusing on any one of them, the orbital magnetic moment is given by
\begin{eqnarray}
	\boldsymbol{m}(\bq)=-\frac{2ev^{2}v_{z}^{2}q_{z}\bq}{\hbar [v^{2}(q_{x}^{2}+q_{y}^{2})+v_{z}^{2}q_{z}^{2}]}.
\end{eqnarray}
Interestingly, the $z$-component of the orbital magnetic momentum on the $q_{z}$ axis is a constant,
\begin{eqnarray}
	m_{z}(0,0,q_{z})=-\frac{2ev^{2}}{\hbar}.\label{OMMz}
\end{eqnarray}
Despite being gapless, an effective mass can be defined for the quadratic band structure,
\begin{eqnarray}
M_{x(y)}^{*}=\hbar^{2}\left(\frac{\partial^{2} E}{\partial q_{x(y)}^{2}}\right)^{-1}=\frac{\hbar^{2}}{2v^{2}}, \quad
M_{z}^{*}=\hbar^{2}\left(\frac{\partial^{2} E}{\partial q_{z}^{2}}\right)^{-1}=\frac{\hbar^{2}}{2v_{z}^{2}}.
\end{eqnarray}
Accordingly, the result in Eq.(\ref{OMMz}) can be reexpressed as
\begin{eqnarray}
m_{z}(0,0,q_{z})=-\frac{e\hbar}{M_{x(y)}^{*}}=-2\frac{e\hbar}{2M_{x(y)}^{*}}=-2\mu_{B}^{*}, \label{Bohr}
\end{eqnarray}
where $\mu_{B}^{*}$ has the interpretation as an effective Bohr magneton. Interestingly, the orbital magnetic moment
is twice of the effective Bohr magneton, which is different from the 2D massive Dirac fermion whose
orbital magnetic moment at the band edge is equal to one effective Bohr magneton~\cite{Xiao2007MM}.


Based on the orbital magnetic moment and Berry curvature, the zero-field orbital magnetization is given by~\cite{Xiao2005OM}
\begin{eqnarray}
	\bM(\mu)=\int\frac{d^{3}q}{(2\pi)^{3}}\Theta(\mu-d(\bq))\left[\boldsymbol{m}(\bq)+\frac{e\bOmega(\bq)}{\hbar}[\mu-d(\bq)]\right].
\end{eqnarray}
Using the relation in Eq.(\ref{OMMBC}), one obtains
\begin{eqnarray}
	\bM(\mu)&=&\int\frac{d^{3}q}{(2\pi)^{3}}\Theta(\mu-d(\bq))\left[\frac{ed(\bq)}{\hbar}\bOmega(\bq)+\frac{e\bOmega}{\hbar}(\mu-d(\bq))\right]\nonumber\\
	&=&\frac{e\mu}{\hbar}\int\frac{d^{3}q}{(2\pi)^{3}}\Theta(\mu-d(\bq))\bOmega(\bq),
\end{eqnarray}
Following the calculations in Eq.(\ref{Hall}), one immediately obtains
\begin{eqnarray}
	\bM(\mu)=-\frac{e}{\hbar}\frac{\mu^{3/2}}{3\pi^{2}v_{z}} \hat{z},\label{Magnetization}
\end{eqnarray}
In the weakly doped regime, for each Berry-dipole node, the contribution to the electron density is
\begin{eqnarray}
n&=&\int\frac{d^{3}q}{(2\pi)^{3}}\Theta(\mu-(v^{2}(q_{x}^{2}+q_{y}^{2})+v_{z}^{2}q_{z}^{2}))\nonumber\\
&=&\frac{1}{v^{2}v_{z}}\int\frac{d^{3}\tilde{q}}{(2\pi)^{3}}\Theta(\mu-\tilde{q}^{2})\nonumber\\
&=&\frac{\mu^{3/2}}{6\pi^{2}v^{2}v_{z}}.
\end{eqnarray}
Using the above result, the orbital magnetization can be reexpressed as
\begin{eqnarray}
	\bM(\mu)=-\frac{2ev^{2}}{\hbar}n \hat{z}=-2n\mu_{B}^{*}\hat{z}.
\end{eqnarray}
Interestingly, the result indicates that all electrons have the same quantized contribution. If the contributions from
two Berry-dipole nodes are both taken into account, the total orbital magnetization is
\begin{eqnarray}
	\bM_{T}(\mu)=2\bM(\mu)=-2n_{T}\mu_{B}^{*}\hat{z}, \label{total}
\end{eqnarray}
where $n_{T}=2n$ is the total electron density. By further using the results in Eq.(\ref{Hall}),
Eq.(\ref{Magnetization}) and Eq.(\ref{total}), a connection between the orbital magnetization and the Hall conductivity
can be built, which reads
\begin{eqnarray}
M_{T,z}(\mu)=\frac{\mu}{e}\sigma_{xy}(\mu).
\end{eqnarray}

\section{III. Anisotropic Landau levels for the low-energy Berry-dipole Hamiltonians}

\subsection{A. Landau levels formed in a $z$-direction magnetic field}

When the energy dispersion is isotropic, usually, the resulting Landau levels will
not rely on the direction of the magnetic field. Two paradigmatic examples
are the Schr\"{o}dinger Hamiltonian $\mathcal{H}(\bq)=\frac{\hbar^{2}q^{2}}{2m}$ and
the Weyl Hamiltonian $\mathcal{H}(\bq)=\hbar v\bq\cdot\boldsymbol{\sigma}$. However,
as will be shown below, even the energy dispersion is isotropic for the low-energy
Berry-dipole Hamiltonian, the Landau levels show remarkable anisotropy when the
magnetic field is applied in different directions.

We first consider that the external magnetic field is applied in the $z$ direction. Focusing
on the orbital effect, the low-energy Berry-dipole Hamiltonian subjected to a $z$-direction magnetic field becomes
\begin{eqnarray}
\mathcal{H}_{\chi}(\bq)=\left(
                   \begin{array}{cc}
                     v^{2}(\Pi_{x}^{2}+\Pi_{y}^{2})-v_{z}^{2}q_{z}^{2} & -2i\chi v v_{z}q_{z}(\Pi_{x}+i\Pi_{y}) \\
                     2i\chi vv_{z}q_{z}(\Pi_{x}-i\Pi_{y}) & -v^{2}(\Pi_{x}^{2}+\Pi_{y})^{2}+v_{z}^{2}q_{z}^{2} \\
                   \end{array}
                 \right),
\end{eqnarray}
where $\Pi_{x}=q_{x}+\frac{eA_{x}}{\hbar}$ and $\Pi_{y}=q_{y}+\frac{eA_{y}}{\hbar}$. To be specific, we
choose the Landau gauge, $\bA=(A_{x},A_{y},A_{z})=B(0,x,0)$. As $[\Pi_{x},\Pi_{y}]=-i\frac{eB}{\hbar}$,
it is convenient to use the Landau ladder operators,
\begin{eqnarray}
\Pi_{x}=\frac{1}{\sqrt{2}l_{B}}(a+a^{\dag}), \quad \Pi_{y}=\frac{i}{\sqrt{2}l_{B}}(a-a^{\dag}),
\end{eqnarray}
where $a$ and $a^{\dag}$ are bosonic annihilation and creation operators satisfying $[a,a^{\dag}]=1$, and
$l_{B}=\sqrt{\frac{\hbar}{eB}}$ is the magnetic length. In terms of the Landau ladder operators, the
Hamiltonian can be rewritten as
\begin{eqnarray}
	\mathcal{H}_{\chi}(q_{z})=\left(\begin{array}{cc}
		\frac{v^{2}}{l_{B}^{2}}(2\hat{n}+1)-v_{z}^{2}q_{z}^{2} & -2\sqrt{2}i\frac{\chi}{l_{B}}vv_{z}q_{z}a^{\dag}\\
		2\sqrt{2}i\frac{\chi}{l_{B}}vv_{z}q_{z}a & -\frac{v^{2}}{l_{B}^{2}}(2\hat{n}+1)+v_{z}^{2}q_{z}^{2}\\
	\end{array}\right),
	\label{eq: Hlow_Bz}
\end{eqnarray}
where $\hat{n}=a^{\dag}a$.

The Landau levels are determined by solving the eigenvalue equation $\mathcal{H}_{\chi}\psi_{\chi}=E_{\chi}\psi_{\chi}$. Choosing
the following trial eigen-functions
\begin{eqnarray}
	\psi_{\chi,n\alpha}=\left(\begin{array}{c}
		u_{\chi,n\alpha}\ket{n}\\
		v_{\chi,n\alpha}\ket{n-1}\\
	\end{array}\right),
\end{eqnarray}
where $\alpha=\pm$, and $\ket{n}$ with the integer $n\geq0$ satisfies $a^{\dag}\ket{n}=\sqrt{n+1}\ket{n+1}$ and $a\ket{n}=\sqrt{n}\ket{n-1}$ for $n\geq1$ and $a\ket{0}=0$.

For $n\geq1$, one can find
\begin{eqnarray} E_{\chi,n\alpha}(q_{z})&=&\frac{v^{2}}{l_{B}^{2}}+\alpha\sqrt{\left(2n\frac{v^{2}}{l_{B}^{2}}-v_{z}^{2}q_{z}^{2}\right)^{2}
+8n\frac{v^{2}v_{z}^{2}q_{z}^{2}}{l_{B}^{2}}},\nonumber\\
	&=&\frac{v^{2}}{l_{B}^{2}}+\alpha(2n\frac{v^{2}}{l_{B}^{2}}+v_{z}^{2}q_{z}^{2}),
	\label{eq: E_Bz_ngeq0}
\end{eqnarray}
and
\begin{eqnarray}	u_{\chi,n\alpha}(q_{z})&=&\sqrt{\frac{1}{2}\left(1+\alpha\frac{2n\frac{v^{2}}{l_{B}^{2}}-v_{z}^{2}q_{z}^{2}}{2n\frac{v^{2}}{l_{B}^{2}}+v_{z}^{2}q_{z}^{2}}\right)},\
	v_{\chi,n\alpha}(q_{z})=i\chi \alpha \sqrt{\frac{1}{2}\left(1-\alpha\frac{2n\frac{v^{2}}{l_{B}^{2}}-v_{z}^{2}q_{z}^{2}}{2n\frac{v^{2}}{l_{B}^{2}}+v_{z}^{2}q_{z}^{2}}\right)},
\end{eqnarray}
For $n=0$, we have
\begin{eqnarray}
	E_{\chi,0}(q_{z})=\frac{\hbar^{2}v^{2}}{l_{B}^{2}}-v_{z}^{2}q_{z}^{2},
	\label{eq: E_Bz_n0}
\end{eqnarray}
and
\begin{eqnarray}
	\psi_{\chi,0}=\left(\begin{array}{c}
		\ket{0}\\
		0\\
	\end{array}\right).
\end{eqnarray}

Three salient properties of the Landau levels are: (i) the Landau levels are quadratic in $q_{z}$ and linear in the field strength
($E\propto l_{B}^{-2}$);
(ii) all Landau levels are separated from each other; (iii) the $0$-th Landau level always cross $E=0$ twice.
The first two properties are similar to the Landau levels for non-relativistic electrons.   The last property
can be understood by noting that the Berry-dipole node corresponds to an overlap of two Weyl nodes with opposite chiralities.

\subsection{B. Landau levels formed in an $x$-direction magnetic field}

Since the low-energy Berry-dipole Hamiltonian has a continuous rotation symmetry about the $q_{z}$ axis, when the magnetic field
is applied in the $xy$ plane, the Landau levels will
not depend on the specific direction of the field in the $xy$ plane. Therefore, we focus on the case that the magnetic
field is applied in the $x$ direction. To simplify the calculation, we first do a unitary transformation to the
low-energy Berry-dipole Hamiltonian,
\begin{eqnarray}
	\mathcal{H}_{\chi}^{'}(\bq)&=&U^{\dagger}\mathcal{H}_{\chi}(\bq)U\nonumber\\
	&=&2\chi vv_{z}q_{y}q_{z}\sigma_{y}+2\chi vv_{z}q_{x}q_{z}\sigma_{z}+\left[v^{2}(q_{x}^{2}+q_{y}^{2})-v_{z}^{2}q_{z}^{2}\right]\sigma_{x},\label{xHamiltonian}
\end{eqnarray}
where the unitary operator reads
\begin{eqnarray}
	U=\frac{1}{\sqrt{2}}\left(\begin{array}{cc}
		1 & 1\\
		i & -i\\
	\end{array}\right).
\end{eqnarray}

After taking the effect of the magnetic field into account, the Hamiltonian becomes
\begin{eqnarray}
	\mathcal{H}_{\chi}^{'}(\bq)
	&=&\chi vv_{z}(\Pi_{y}\Pi_{z}+\Pi_{z}\Pi_{y})\sigma_{y}+2\chi vv_{z}q_{x}\Pi_{z}\sigma_{z}+\left[v^{2}(q_{x}^{2}+\Pi_{y}^{2})-v_{z}^{2}\Pi_{z}^{2}\right]\sigma_{x},\label{xfield}
\end{eqnarray}
where $\Pi_{y}=q_{y}+\frac{eA_{y}}{\hbar}$ and $\Pi_{z}=q_{z}+\frac{eA_{z}}{\hbar}$. To be specific, we
again choose the Landau gauge, $\bA=(0,A_{y},A_{z})=B(0,0,y)$. It is worth noting that, since
$\Pi_{y}$ and $\Pi_{z}$ do not commute, we have adopted symmetrization when the two operators
show together, i.e., $2\Pi_{y}\Pi_{z}\rightarrow(\Pi_{y}\Pi_{z}+\Pi_{z}\Pi_{y})$.  Similar to
the previous case, it is convenient to use the Landau ladder operators,
\begin{eqnarray}
	\Pi_{y}=\sqrt{\frac{v_{z}}{2v}}\frac{1}{l_{B}}(a+a^{\dagger}), \Pi_{z}=i\sqrt{\frac{v}{2v_{z}}}\frac{1}{l_{B}}(a-a^{\dagger}).
\end{eqnarray}
Here we have performed a rescaling of the operators to eliminate the anisotropy of the velocity in the $yz$ plane.
Taking the above expressions back into Eq.(\ref{xfield}), one obtains
\begin{eqnarray}
\mathcal{H}_{\chi}^{'}(\bq)
	&=&i\chi \frac{vv_{z}}{l_{B}^{2}}[a^{2}-(a^\dag)^{2}]\sigma_{y}+i\chi\frac{\sqrt{2 v^{3}v_{z}}}{l_{B}}q_{x}(a-a^{\dag})\sigma_{z}+\left[v^{2}q_{x}^{2}+\frac{vv_{z}}{l_{B}^{2}}[a^{2}+(a^{\dag})^{2}]\right]\sigma_{x},
\end{eqnarray}
The exact Landau levels for the above Hamiltonian cannot be analytically obtained. Below we adopt a
perturbation approach to approximately determine the analytical form of the Landau levels.  Concretely, we divide the
Hamiltonian into two parts,
\begin{eqnarray}
	\mathcal{H}_{\chi}^{'}(q_{x})&=&\mathcal{H}_{\chi}^{'(0)}+\mathcal{H}_{\chi}^{'(1)}(q_{x}),
	\label{eq: Hlow_Bx}
\end{eqnarray}
where
\begin{eqnarray}
	\mathcal{H}_{\chi}^{'(0)}&=&i\chi\frac{vv_{z}}{l_{B}^{2}}\left[a^{2}-(a^{\dagger})^{2}\right]\sigma_{y}+\frac{vv_{z}}{l_{B}^{2}}\left[a^{2}+(a^{\dagger})^{2}\right]\sigma_{x}\nonumber\\
	&=&\left(\begin{array}{cc}
		0 & \frac{vv_{z}}{l_{B}^{2}}\left[a^{2}(1+\chi)+(a^{\dagger})^{2}(1-\chi)\right]\\
		\frac{vv_{z}}{l_{B}^{2}}\left[a^{2}(1-\chi)+(a^{\dagger})^{2}(1+\chi)\right] & 0\\
	\end{array}\right),
	\label{eq: Hlow_Bx0}
\end{eqnarray}
and
\begin{eqnarray}
	\mathcal{H}_{\chi}^{'(1)}(q_{x})&=&v^{2}q_{x}^{2}\sigma_{x}+i\chi\frac{\sqrt{2v^{3}v_{z}}}{l_{B}}q_{x}(a-a^{\dagger})\sigma_{z}.
\end{eqnarray}
As $\mathcal{H}_{\chi}^{'(0)}$ has no dependence on $q_{x}$ and $\mathcal{H}_{\chi}^{'(1)}(q_{x})$ depends
on $q_{x}$,  it is justified to treat $\mathcal{H}_{\chi}^{'(1)}$ as a perturbation when $q_{x}$ is small,
the regime that we are mostly interested in.

As the Hamiltonians for the two Berry-dipole nodes are different, we analyze them separately.
Let us first study the case with $\chi=1$.  The Hamiltonian (\ref{eq: Hlow_Bx0}) for this case reduces as
\begin{eqnarray}
	\mathcal{H}_{\chi=1}^{'(0)}
	&=&\frac{2vv_{z}}{l_{B}^{2}}\left(\begin{array}{cc}
		0 & a^{2}\\
		(a^{\dagger})^{2}& 0\\
	\end{array}\right).
\end{eqnarray}
$\mathcal{H}_{\chi=1}^{'(0)}$ is analytically solvable. Adopting the trial wave function,
\begin{eqnarray}
	\psi_{n\alpha}^{(0)}=\left(\begin{array}{c}
		u_{n\alpha}\ket{n-2}\\
		v_{n\alpha}\ket{n}\\
	\end{array}\right),
\end{eqnarray}
and solving the eigenvalue equation $\mathcal{H}_{\chi=1}^{'(0)}\psi_{n\alpha}^{(0)}=\epsilon_{n\alpha}\psi_{n\alpha}^{(0)}$,
we find that, for $n\geq2$,
\begin{eqnarray}
	\epsilon_{n\alpha}=\alpha\frac{2vv_{z}}{l_{B}^{2}}\sqrt{n(n-1)},\quad \psi_{n\alpha}^{(0)}=\frac{1}{\sqrt{2}}\left(\begin{array}{c}
		\alpha\ket{n-2}\\
		\ket{n}\\
	\end{array}\right),
\end{eqnarray}
where $\alpha=\pm$. For $n=0$ and $n=1$, there are two degenerate Landau levels at zero energy, i.e. $\epsilon_{0}=\epsilon_{1}=0$,
with their wave functions given by
\begin{eqnarray}
	\psi_{0}^{(0)}=\left(\begin{array}{c}
		0\\
		\ket{0}\\
	\end{array}\right),\
	\psi_{1}^{(0)}=\left(\begin{array}{c}
		0\\
		\ket{1}\\
	\end{array}\right).
\end{eqnarray}
These two degenerate Landau levels at zero energy can be understood by noting  that the low-energy Berry-dipole Hamiltonian
at the high-symmetry momentum plane $q_{x}=0$ reduces to a quadratic Dirac Hamiltonian which has been
extensively studied in the context of bilayer graphene~\cite{Novoselov2006}. The quadratic Dirac Hamiltonian has the property that the Berry phase along a closed path around the node is $2\pi$, or equivalently, the winding number is $2$ as the reduced Hamiltonian has chiral symmetry.
The winding number can protect two robust zero-energy states.

Based on the wave functions of the analytically solvable part, now we determine the
correction from the perturbation part. First, we project $\mathcal{H}_{\chi=1}^{'(1)}(q_{x})$ onto
the basis spanned by $\{\psi_{n\alpha}^{(0)}\}$ and obtain the matrix elements,
\begin{eqnarray}
\mathcal{H}_{\chi=1,(m\alpha,n\beta)}^{'(1)}(q_{x})=(\psi_{m\alpha}^{(0)})^{\dagger}\mathcal{H}_{\chi=1}^{'(1)}(q_{x})\psi_{n\beta}^{(0)}.
\end{eqnarray}
By a straightforward calculation,
one can find that the
nonzero elements are
$\mathcal{H}_{\chi=1,(0,1)}^{'(1)}(q_{x})=-i \sqrt{2v^{3}v_{z}}q_{x}/l_{B}$,
$\mathcal{H}_{\chi=1,(0,2\pm)}^{'(1)}(q_{x})=\mathcal{H}_{\chi=1,(1,3\pm)}^{'(1)}(q_{x})=\pm v^{2}q_{x}^{2}/\sqrt{2}$,
$\mathcal{H}_{\chi=1,(1,2\pm)}^{'(1)}(q_{x})=-i\sqrt{2v^{3}v_{z}}q_{x}/l_{B}$,
and for $n\geq2$,
\begin{eqnarray}
\mathcal{H}_{\chi=1,[n\alpha,(n+1)\beta]}^{'(1)}(q_{x})&=&-i\frac{\sqrt{2v^{3}v_{z}}}{2l_{B}}q_{x}(\sqrt{n+1}
-\alpha\beta\sqrt{n-1}),\nonumber\\
\mathcal{H}_{\chi=1,[(n+1)\alpha,n\beta]}^{'(1)}(q_{x})&=&i\frac{\sqrt{2v^{3}v_{z}}}{2l_{B}}q_{x}(\sqrt{n+1}
-\alpha\beta\sqrt{n-1}),\nonumber\\
\mathcal{H}_{\chi=1,[n\alpha,(n+2)\beta]}^{'(1)}(q_{x})&=&\beta v^{2}q_{x}^{2}/2,\nonumber\\
\mathcal{H}_{\chi=1,[(n+2)\alpha,n\beta]}^{'(1)}(q_{x})&=&\alpha v^{2}q_{x}^{2}/2.
\end{eqnarray}
For notational simplicity, we define $A_{1}=-i \sqrt{2v^{3}v_{z}}q_{x}/l_{B}$, $B=v^{2}q_{x}^{2}/\sqrt{2}$,
and $A_{n\pm}=-i\sqrt{2v^{3}v_{z}}q_{x}(\sqrt{n+1}
\pm\sqrt{n-1})/(2l_{B})$ for $n\geq2$. If one
chooses the basis to be $\{\psi_{0}^{(0)},\psi_{1}^{(0)},\psi_{2+}^{(0)},\psi_{2-}^{(0)},...,\psi_{n+}^{(0)},\psi_{n-}^{(0)},...\}$,
one can find that the corresponding matrix form of $\mathcal{H}_{\chi=1}^{'(1)}(q_{x})$ is
\begin{eqnarray}
\mathcal{H}_{\chi=1}^{'(1)}(q_{x})=\left(
                                     \begin{array}{ccccccccccc}
                                       0 & A_{1} & B & -B & 0 & 0 & 0 & 0 & 0 & 0 & \cdots \\
                                       A_{1}^{*} & 0 & A_{1} & A_{1} & B & -B & 0 & 0 & 0 & 0 & \cdots \\
                                       B & A_{1}^{*} & 0 & 0 & A_{2-} & A_{2+} & \frac{B}{\sqrt{2}} & -\frac{B}{\sqrt{2}} & 0 & 0 & \cdots \\
                                       -B & A_{1}^{*} & 0 & 0 & A_{2+} & A_{2-} & \frac{B}{\sqrt{2}} & -\frac{B}{\sqrt{2}} & 0 & 0 & \cdots \\
                                       0 & B & A_{2-}^{*} & A_{2+}^{*} & 0 & 0 & A_{3-} & A_{3+} & \frac{B}{\sqrt{2}} & -\frac{B}{\sqrt{2}} & \cdots \\
                                       0 & -B & A_{2+}^{*} & A_{2-}^{*} & 0 & 0 & A_{3+} & A_{3-} & \frac{B}{\sqrt{2}} & -\frac{B}{\sqrt{2}} & \cdots \\
                                       0 & 0 & \frac{B}{\sqrt{2}} & \frac{B}{\sqrt{2}} & A_{3-}^{*} & A_{3+}^{*} & 0 & 0 &  A_{4-} & A_{4+} & \cdots \\
                                       0 & 0 & -\frac{B}{\sqrt{2}} & -\frac{B}{\sqrt{2}} & A_{3+}^{*} & A_{3-}^{*} & 0 & 0 & A_{4+} &  A_{4-} & \cdots \\
                                       0 & 0 & 0 & 0 & \frac{B}{\sqrt{2}} & \frac{B}{\sqrt{2}} & A_{4-}^{*} & A_{4+}^{*} & 0 & 0 & \cdots \\
                                       0 & 0 & 0 & 0 & -\frac{B}{\sqrt{2}} & -\frac{B}{\sqrt{2}} & A_{4+}^{*} & A_{4-}^{*} & 0 & 0 & \cdots \\
                                       \vdots & \vdots & \vdots & \vdots & \vdots & \vdots & \vdots & \vdots & \vdots & \vdots & \ddots \\
                                     \end{array}
                                   \right).
\end{eqnarray}
In this basis, $\mathcal{H}_{\chi=1}^{'(0)}$ is diagonal, i.e., $\mathcal{H}_{\chi=1}^{'(0)}=\text{diag}\{0,0,\epsilon_{2+},\epsilon_{2-},...,\epsilon_{n+},\epsilon_{n-},...\}$.  Adding $\mathcal{H}_{\chi=1}^{'(0)}$ and $\mathcal{H}_{\chi=1}^{'(1)}(q_{x})$ together, one obtains the total Hamiltonian, and  then a direct
diagonalization of the total Hamiltonian can numerically determine the exact form of the Landau levels.

In the following, we use  perturbation theory to obtain an approximate but analytical expression of the Landau levels.
Since there are two degenerate Landau levels at $q_{x}=0$, we first perform a unitary transformation
to the Hamiltonian. The unitary matrix has the form
\begin{eqnarray}
U=\left(
    \begin{array}{ccccccc}
      \frac{\sqrt{2}}{2} & \frac{\sqrt{2}i}{2} & 0 & 0 & 0 & 0 & \cdots \\
      \frac{\sqrt{2}i}{2} & \frac{\sqrt{2}}{2} & 0 & 0 & 0 & 0 & \cdots \\
      0 & 0 & 1 & 0 & 0 & 0 & \cdots \\
      0 & 0 & 0 & 1 & 0 & 0 & \cdots \\
      0 & 0 & 0 & 0 & 1 & 0 & \cdots \\
      0 & 0 & 0 & 0 & 0 & 1 & \cdots \\
      \vdots & \vdots & \vdots & \vdots & \vdots & \vdots & \ddots \\
    \end{array}
  \right).
\end{eqnarray}
The action of $U$ on $\mathcal{H}_{\chi=1}^{'(0)}$ does not change it, i.e.,
$\tilde{\mathcal{H}}_{\chi=1}^{'(0)}=U^{\dag}\mathcal{H}_{\chi=1}^{'(0)}U=\mathcal{H}_{\chi=1}^{'(0)}$.
While the action of  $U$ on $\mathcal{H}_{\chi=1}^{'(1)}(q_{x})$ leads to
\begin{eqnarray}
\tilde{\mathcal{H}}_{\chi=1}^{'(1)}(q_{x})&=&U^{\dag}\mathcal{H}_{\chi=1}^{'(1)}(q_{x})U\nonumber\\
&=&\left(
                                     \begin{array}{ccccccccccc}
                                       \epsilon_{1+} & 0 & C_{-} & -C_{+} & -\frac{\sqrt{2}i}{2}B & \frac{\sqrt{2}i}{2}B & 0 & 0 & 0 & 0 & \cdots \\
                                       0 & \epsilon_{1-} & -iC_{+} & iC_{-} & \frac{\sqrt{2}}{2}B & -\frac{\sqrt{2}}{2}B & 0 & 0 & 0 & 0 & \cdots \\
                                       C_{-} & iC_{+} & 0 & 0 & A_{2-} & A_{2+} & \frac{B}{\sqrt{2}} & -\frac{B}{\sqrt{2}} & 0 & 0 & \cdots \\
                                       -C_{+} & -iC_{-} & 0 & 0 & A_{2+} & A_{2-} & \frac{B}{\sqrt{2}} & -\frac{B}{\sqrt{2}} & 0 & 0 & \cdots \\
                                       \frac{\sqrt{2}i}{2}B & \frac{\sqrt{2}}{2}B & A_{2-}^{*} & A_{2+}^{*} & 0 & 0 & A_{3-} & A_{3+} & \frac{B}{\sqrt{2}} & -\frac{B}{\sqrt{2}} & \cdots \\
                                       -\frac{\sqrt{2}i}{2}B & -\frac{\sqrt{2}}{2}B & A_{2+}^{*} & A_{2-}^{*} & 0 & 0 & A_{3+} & A_{3-} & \frac{B}{\sqrt{2}} & -\frac{B}{\sqrt{2}} & \cdots \\
                                       0 & 0 & \frac{B}{\sqrt{2}} & \frac{B}{\sqrt{2}} & A_{3-}^{*} & A_{3+}^{*} & 0 & 0 &  A_{4-} & A_{4+} & \cdots \\
                                       0 & 0 & -\frac{B}{\sqrt{2}} & -\frac{B}{\sqrt{2}} & A_{3+}^{*} & A_{3-}^{*} & 0 & 0 & A_{4+} &  A_{4-} & \cdots \\
                                       0 & 0 & 0 & 0 & \frac{B}{\sqrt{2}} & \frac{B}{\sqrt{2}} & A_{4-}^{*} & A_{4+}^{*} & 0 & 0 & \cdots \\
                                       0 & 0 & 0 & 0 & -\frac{B}{\sqrt{2}} & -\frac{B}{\sqrt{2}} & A_{4+}^{*} & A_{4-}^{*} & 0 & 0 & \cdots \\
                                       \vdots & \vdots & \vdots & \vdots & \vdots & \vdots & \vdots & \vdots & \vdots & \vdots & \ddots \\
                                     \end{array}
                                   \right).
\end{eqnarray}
where $\epsilon_{1\pm}=\pm\sqrt{2v^{3}v_{z}}q_{x}/l_{B}$, and $C_{\pm}=\sqrt{2}(B\pm iA_{1})/2$.
After this operation, we can rearrange the Hamiltonian into two parts, i.e., $\tilde{\mathcal{H}}_{\chi=1}'=\mathcal{H}_{d}+\mathcal{H}_{p}$, where
\begin{eqnarray}
\mathcal{H}_{d}&=&\text{diag}\{\epsilon_{1},\epsilon_{1-},\epsilon_{2+},\epsilon_{2-},...,\epsilon_{n+},\epsilon_{n-},...\},\nonumber\\
\mathcal{H}_{p}&=&\left(
                                     \begin{array}{ccccccccccc}
                                        0 & 0 & C_{-} & -C_{+} & -\frac{\sqrt{2}i}{2}B & \frac{\sqrt{2}i}{2}B & 0 & 0 & 0 & 0 & \cdots \\
                                       0 & 0 & -iC_{+} & iC_{-} & \frac{\sqrt{2}}{2}B & -\frac{\sqrt{2}}{2}B & 0 & 0 & 0 & 0 & \cdots \\
                                       C_{-} & iC_{+} & 0 & 0 & A_{2-} & A_{2+} & \frac{B}{\sqrt{2}} & -\frac{B}{\sqrt{2}} & 0 & 0 & \cdots \\
                                       -C_{+} & -iC_{-} & 0 & 0 & A_{2+} & A_{2-} & \frac{B}{\sqrt{2}} & -\frac{B}{\sqrt{2}} & 0 & 0 & \cdots \\
                                       \frac{\sqrt{2}i}{2}B & \frac{\sqrt{2}}{2}B & A_{2-}^{*} & A_{2+}^{*} & 0 & 0 & A_{3-} & A_{3+} & \frac{B}{\sqrt{2}} & -\frac{B}{\sqrt{2}} & \cdots \\
                                       -\frac{\sqrt{2}i}{2}B & -\frac{\sqrt{2}}{2}B & A_{2+}^{*} & A_{2-}^{*} & 0 & 0 & A_{3+} & A_{3-} & \frac{B}{\sqrt{2}} & -\frac{B}{\sqrt{2}} & \cdots \\
                                       0 & 0 & \frac{B}{\sqrt{2}} & \frac{B}{\sqrt{2}} & A_{3-}^{*} & A_{3+}^{*} & 0 & 0 &  A_{4-} & A_{4+} & \cdots \\
                                       0 & 0 & -\frac{B}{\sqrt{2}} & -\frac{B}{\sqrt{2}} & A_{3+}^{*} & A_{3-}^{*} & 0 & 0 & A_{4+} &  A_{4-} & \cdots \\
                                       0 & 0 & 0 & 0 & \frac{B}{\sqrt{2}} & \frac{B}{\sqrt{2}} & A_{4-}^{*} & A_{4+}^{*} & 0 & 0 & \cdots \\
                                       0 & 0 & 0 & 0 & -\frac{B}{\sqrt{2}} & -\frac{B}{\sqrt{2}} & A_{4+}^{*} & A_{4-}^{*} & 0 & 0 & \cdots \\
                                       \vdots & \vdots & \vdots & \vdots & \vdots & \vdots & \vdots & \vdots & \vdots & \vdots & \ddots \\
                                     \end{array}
                                   \right).
\end{eqnarray}
Now we use perturbation theory to obtain the correction from $\mathcal{H}_{p}$ to the energy levels of $\mathcal{H}_{d}$.
The result is
\begin{eqnarray}
E_{n\alpha}=\epsilon_{n\alpha}+\sum_{m\neq n,\beta}\frac{|(H_{p})_{n\alpha,m\beta}|^{2}}{\epsilon_{n\alpha}-\epsilon_{m\beta}},
\end{eqnarray}
For the four Landau levels most close to $E=0$, their explicit expressions are
\begin{eqnarray}
E_{1+}(q_{x})&=&-E_{1-}(q_{x})=\frac{\sqrt{2v^{3}v_{z}}}{l_{B}}q_{x}+\frac{|C_{-}|^{2}}{\frac{\sqrt{2v^{3}v_{z}}}{l_{B}}q_{x}-\frac{2\sqrt{2}vv_{z}}{l_{B}^{2}}}
+\frac{|C_{+}|^{2}}{\frac{\sqrt{2v^{3}v_{z}}}{l_{B}}q_{x}+\frac{2\sqrt{2}vv_{z}}{l_{B}^{2}}}\nonumber\\
&&+\frac{B^{2}}{2[\frac{\sqrt{2v^{3}v_{z}}}{l_{B}}q_{x}-\frac{2\sqrt{6}vv_{z}}{l_{B}^{2}}]}
+\frac{B^{2}}{2[\frac{\sqrt{2v^{3}v_{z}}}{l_{B}}q_{x}+\frac{2\sqrt{6}vv_{z}}{l_{B}^{2}}]},
\end{eqnarray}
and
\begin{eqnarray}
E_{2+}(q_{x})&=&-E_{2-}(q_{x})=\frac{2\sqrt{2}vv_{z}}{l_{B}^{2}}+\frac{|C_{-}|^{2}}{\frac{2\sqrt{2}vv_{z}}{l_{B}^{2}}-\frac{\sqrt{2v^{3}v_{z}}}{l_{B}}q_{x}}
+\frac{|C_{+}|^{2}}{\frac{2\sqrt{2}vv_{z}}{l_{B}^{2}}+\frac{\sqrt{2v^{3}v_{z}}}{l_{B}}q_{x}}\nonumber\\
&&+\frac{|A_{2-}|^{2}}{[\frac{2\sqrt{2}vv_{z}}{l_{B}^{2}}-\frac{2\sqrt{6}vv_{z}}{l_{B}^{2}}]}
+\frac{|A_{2+}|^{2}}{[\frac{2\sqrt{2}vv_{z}}{l_{B}^{2}}+\frac{2\sqrt{6}vv_{z}}{l_{B}^{2}}]}\nonumber\\
&&+\frac{B^{2}}{2[\frac{2\sqrt{2}vv_{z}}{l_{B}^{2}}-\frac{2\sqrt{12}vv_{z}}{l_{B}^{2}}]}
+\frac{B^{2}}{2[\frac{2\sqrt{2}vv_{z}}{l_{B}^{2}}+\frac{2\sqrt{12}vv_{z}}{l_{B}^{2}}]}.
\end{eqnarray}
If only the leading order term in momentum is kept, one obtains the following compact expressions of
the Landau levels,
\begin{eqnarray}
E_{1\pm}(q_{x})&=&\pm\frac{\sqrt{2v^{3}v_{z}}}{l_{B}}q_{x},\nonumber\\
E_{n\pm}(q_{x})&=&\pm[\frac{2\sqrt{n(n-1)}vv_{z}}{l_{B}^{2}}+\frac{(2n-1)}{2\sqrt{n(n-1)}}v^{2}q_{x}^{2}], \, n\geq2.
\end{eqnarray}
Notably, the above results indicate that the two Landau levels most close to zero energy
are of linear dispersion and cross at zero energy, which is fundamentally different from
the gapped behavior of the Landau levels when the magnetic field is applied in the $z$ direction.
Furthermore, here the Landau levels are symmetric about $E=0$, which is also different
from the Landau levels when the magnetic field is applied in the $z$-direction.
Interestingly, here the crossing characteristic of the two linear Landau levels
is reminiscent of the two linear zeroth
Landau levels of a 3D Dirac Hamiltonian. However,
a big difference is that the two linear zeroth Landau levels associated with a Dirac node
are of opposite spin polarizations, in contrast, here the two linear Landau levels associated
with a Berry-dipole node are of the same spin polarization.

Similar steps can also obtain the Landau levels for the other Berry-dipole Hamiltonian. The Landau levels are of the same
structure. The main difference is that the Landau levels will have different spin polarizations.

\end{widetext}

\end{document}